%% file: aphot.tex
\documentclass[iop]{emulateapj}
\usepackage{epstopdf}
\usepackage{natbib}
\usepackage{amsmath}
\usepackage{enumitem}
\setlist[enumerate]{itemsep=0mm}
\usepackage{url}

\usepackage{color}
\usepackage[usenames,dvipsnames,svgnames,table]{xcolor}

\DeclareGraphicsExtensions{.jpg,.pdf,.png,.eps,.ps} \graphicspath{. }

\shortauthors{Nurgaliev et~al.}
\shorttitle{Galaxy Cluster Morphology Using Asymmetry and Central Concentration}

\definecolor{darkblue}{rgb}{0,0,0.9}

\begin{document}

\newcommand{\kpc}{\mathrm{kpc}}
\newcommand{\keV}{\mathrm{keV}}
\newcommand{\Flux}{\mathrm{Flux}}
\newcommand{\new}{\mathrm{new}}
\newcommand{\old}{\mathrm{old}}
\providecommand{\e}[1]{\ensuremath{\times 10^{#1}}}

\title{Robust Quantification of Galaxy Cluster Morphology Using Asymmetry and 
Central Concentration}

\input{six_authors}
\email{nurgaliev@physics.harvard.edu}
\begin{abstract}

  We present a novel quantitative scheme of cluster classification
  based on the morphological properties that are manifested in X-ray
  images. We use a conventional radial surface brightness
  concentration parameter ($\rm{c_{SB}}$) as defined previously by
  others, and a new asymmetry parameter, which we define in this
  paper. Our asymmetry parameter, which we refer to as \emph{photon
    asymmetry} ($\rm{A_{phot}}$), was developed as a robust
  substructure statistic for cluster observations with only a few
  thousand counts.  To demonstrate that photon asymmetry exhibits
  better stability than currently popular power ratios and centroid
  shifts, we artificially degrade the X-ray image quality by: (a)
  adding extra background counts, (b) eliminating a fraction of the
  counts, (c) increasing the width of the smoothing kernel, and (d)
  simulating cluster observations at higher redshift. The
    asymmetry statistic presented here has a smaller statistical
    uncertainty than competing substructure parameters, allowing for
    low levels of substructure to be measured with
    confidence. $\rm{A_{phot}}$ is less sensitive to the total number
    of counts than competing substructure statistics, making it an
    ideal candidate for quantifying substructure in samples of distant
    clusters covering wide range of observational S/N.  Additionally,
    we show that the asymmetry-concentration classification separates
    relaxed, cool core clusters from morphologically-disturbed
    mergers, in agreement with by-eye classifications.  Our algorithms, freely
  available as \emph{Python} scripts
  (\url{\texttt{https://github.com/ndaniyar/aphot}}) are completely automatic 
  and can be used to rapidly classify galaxy cluster morphology for large
  numbers of clusters without human intervention.  
  
\end{abstract}
\smallskip

\section{Introduction}

Clusters of galaxies are complex objects where many astrophysical processes are 
taking place. Cluster classification based on their X-ray morphology can help us 
understand the dominant physical processes in particular types of clusters, shed 
light on their formation histories, and give new insights into the evolution of 
both the large scale structure of the Universe \citep{Allen2011} and the 
baryonic component of galaxy clusters \citep{Bohringer2010}.

Two distinctive features of galaxy clusters that are detectable in X-ray images 
are 1) cool cores and 2) departure from axial symmetry, presumed to arise from 
galaxy cluster mergers. Cool cores exhibit sharp central peaks in X-ray 
emission, while asymmetry manifests as secondary peaks, filaments, and clumps in 
X-ray surface brightness. It is believed that these features emerge at different 
stages of cluster evolution, and are outcomes of completely different physical 
processes that affect the entire intracluster medium (ICM). One important reason 
to classify cluster morphology is that we can explore any correlations between 
morphology and residuals in various cluster scaling relations, resulting in more 
robust estimates of, for example, galaxy cluster mass (M$_{500}$).

 The substructure clumps in the X-ray emission are often associated with active 
processes of dynamical relaxation after mergers. For such clusters (with a high 
amount of substructure) the characteristic processes are turbulence 
\citep{Vazza2011, Hallman2011}, shocks, and cold fronts in the ICM 
\citep{Markevitch2007, Hallman2010, Blanton2011}, giant and mini radio halos 
\citep{Cassano2010} and relics \citep{Ferrari2008}.  After the process of 
relaxation is over, cool cores start to develop 
\citep{Fabian1994,Peterson2006,Hudson2010,McDonald2013}, and the evolution of 
the ICM is governed by the processes of gas cooling and heating, AGN feedback 
\citep{McNamara2007} and thermal conduction \citep{Voit2011}. We are still far 
from a detailed understanding of these processes, but their correlation with 
morphology is established both from observations and simulations. For example, 
observations suggest that more dynamically disturbed systems have weaker cool 
cores \citep{Sanderson2009}.

In this work, we propose a new classification scheme, based on the arrangement 
of galaxy clusters in the 2-dimensional plane of \emph{disturbance} and 
\emph{cool core strength}. As explained above, this choice of fundamental 
morphological parameters is observationally well-motivated. To choose the 
parameters that best quantify cool core strength and disturbance, we first 
formulate some requirements:
\begin{enumerate}
  \item These parameters need to be objective, quantitative and reproducible.
 
  \item The parameters should be model independent.
 
  \item They should allow substructure analysis for low signal-to-noise (S/N) 
 observations. 
 
  \item These parameters should be relatively insensitive to exposure time, the 
level of the X-ray background or a cluster's angular size on the sky. A 
composite test that checks all these sensitivities together is simulating 
observations of a cluster at higher redshift.
 
  \item The substructure parameters should agree with the human expert 
      judgement.
\end{enumerate}

 The radial surface brightness profile of X-ray emission can be used to quantify 
the extent to which a cool core is present, although assigning clusters to 
categories (cool core vs. non cool core) is still a topic of discussion 
\citep{Hudson2010,McDonald2013}. We adopt here the concentration prescription of 
\citet{Santos2008}, who showed that their implementation can discriminate 
between ``strong'', ``medium'' and ``no'' cool cores. Importantly, in the 
context of the requirements listed earlier, the \citet{Santos2008} concentration 
parametrization is robust even for low S/N observations and is roughly model 
independent.
 
The quantification of ``disturbance'' is significantly harder. There
is no simple physical (or mathematical) quantity that can measure
``disturbance'' or as it is usually called, the amount of
substructure. \citet{Pinkney1996} list 30 different substructure tests
and conclude that no single one is good in all cases. Two substructure
statistics have, nevertheless, became popular recently: centroid
shifts \citep{Mohr1993} and power ratios \citep{Buote&Tsai1995,
  Buote&Tsai1996}. Their popularity can be explained by their
model-independence and ease of computation.  They also satisfy,
reasonably well, the requirements formulated above. For a more
detailed review of various substructure statistics see \citet{Buote2002,
  Bohringer2010a, Rasia2013, Weissmann2013}. We present below a new
substructure statistic that is superior based on the above
  requirements.

We stress that any substructure statistic should be suitable for high redshift 
clusters, with observations of poor quality. This is an area where the other 
substructure tests do not perform very well. Most morphological studies have 
been carried out for nearby clusters with high S/N X-ray images ($10^5$ counts 
per cluster being the typical value for these studies). However, large surveys 
or serendipitous discoveries of high-redshift clusters will yield images with, 
typically, only several hundred counts \citep[e.g.,][]{McDonald2013}.  Thus, a 
reliable next-generation substructure statistic must perform equally well on 
low-S/N, high-redshift
observations.

Here, we present a new substructure statistic, \emph{photon asymmetry}  
($\rm{A_{phot}}$), which quantifies how much the X-ray emission deviates from 
the idealized axisymmetric case. This statistic is somewhat similar to existing 
efforts to use the residuals after subtracting a beta-model fit 
\citep[e.g.][]{Neumann1997, Bohringer2000, Andrade-Santos2012} or double 
beta-model \citep{Mohr1999}. However, $\rm{A_{phot}}$ is model independent and 
specially designed to work well for observations with low photon counts.

In \S2, we present the X-ray sample that has been used to develop our approach. \S3 
defines the various morphology measures that are compared in this work, while \S4 explores 
performance using simulated data sets. Our results and conclusions are presented in 
\S5 and \S6, respectively. We defer an analysis of how these morphological 
parameters correlate with the scaling relations residuals to a future 
publication.

\bigskip
\section{Sample and data reduction}

\subsection{Sample}

 To test our classification method and compare the properties of photon 
asymmetry to the properties of previously known substructure statistics, we used 
the high-z subsample of the 400 square degree galaxy cluster survey (abbreviated 
as 400d), which is a quasi mass-limited sample of galaxy clusters at $z>0.35$ 
serendipitously detected in ROSAT PSPC data \citep{Burenin2007}.
 The high-redshift subsample of 400d was published in \citet{CCCPII} and 
consists of 36 clusters with $z > 0.35$ which exceed a certain luminosity 
threshold  which corresponds to $\approx 10^{14} M_{\Sun}$ in mass (see 
\citet{CCCPII} for details).

All clusters in the sample have been observed with the \emph{Chandra X-ray 
Observatory} and used to constrain cosmological parameters in \citet{CCCPII, 
CCCPIII}.

The reasons for choosing this cluster sample are:
\begin{enumerate}
    \item A redshift range that covers $0.3 \le z \le 0.9$, and is similar to 
the redshift range of both SZ surveys and next-generation X-ray surveys (e.g., 
eRosita), allowing extension to larger samples in the future.

    \item High-resolution Chandra imaging which is very suitable for 
substructure detection. As we show in the paper, telescope resolution is very 
important to detect and quantify substructure.

    \item A range of photon counts. Since our goal is to develop a substructure 
statistic that is maximally applicable to high-z clusters with low S/N 
observations, the high-z part of the 400d catalog is perfect for testing our 
substructure statistic.  

    \item The basic selection criterion is X-ray luminosity, which adequately 
samples the range of cluster morphologies and core properties. Thus the sample 
should be representative with respect to cluster morphological types.

\end{enumerate}

\subsection{Data reduction}

We perform all industry-standard X-ray data reduction steps. We start with flare 
 cleaned event2 files that are identical to those used by \cite{CCCPII}.
Following many other cluster studies \citep[e.g.][]{Santos2008}, we apply a 0.5 
- 5.0 keV band filter which optimizes the ratio of the cluster to background 
flux.
 We chose to use a higher upper cut-off than what was used in many other studies 
(2 keV), because for massive clusters there is significant emission above 2 keV.

 We detect point sources with an algorithm similar to \emph{wavdetect} from the CIAO 
package \citep{CIAO} and replace the regions of point sources with a Poisson 
distribution with a mean value equal to the local background density of counts.  
In most cases this means that we add no counts in the region of the removed 
point source because the typical local background level is  $\sim 10^{-2}$ 
counts per pixel.

We estimate the global background level from regions on the chip free of point sources,
away from chip gaps, and sufficiently far away from cluster center (2-4 
R$_{500}$ annulus)

 We compute all morphological parameters directly from the raw event2 
band-filtered files without additional binning or smoothing.
All substructure statistics that we consider in this paper can be formulated in 
terms of sums over counts instead of integrals over surface brightness 
distributions as they are usually presented.
 We believe that this is the best way to perform statistical tests because any 
post-processing may distort and bias the statistic's distributions.

 We use exposure maps that include corrections for CCD gaps, spatial variations 
of the effective area, ACIS contamination, bad pixels and detector quantum 
efficiency.

We produce smoothed images of the clusters using an algorithm similar to 
\emph{asmooth} \citep{Ebeling2006}, which chooses the appropriate smoothing 
scale adaptively for each count based on the local density of counts. These 
smoothed images are used for two (and only two) purposes:
\begin{enumerate}
    \item Visualization for by-eye classification and by-eye comparison of the 
         cluster's relative ranking produced by various substructure statistics, 
     \item Generation of simulated cluster observations. See 
         Section~\ref{sec:uncertainties} for more details.
\end{enumerate}

All the steps in the data reduction pipeline are automatic, but the results of 
each step were visually inspected. For the clusters that had several 
observations, we merged all observations that had the entire $R_{500}$ aperture on the CCD.

\bigskip
\section{Classical morphological parameters/ substructure statistics}
\label{sec:morph_params}

\subsection{Power ratios}

Power ratios were introduced in \citet{Buote&Tsai1995, Buote&Tsai1996} and have 
been widely used ever since \citep[e.g.][]{Jeltema2005, Ventimiglia2008, 
Cassano2010}.
 They are able to distinguish a large range of morphologies, physically 
 motivated and easy to compute \citep{Jeltema2005}.
 The method consists of a multipole expansion of the surface brightness  and 
computes the powers in different orders of the expansion.
 The corresponding formulas are usually quoted as integrals over surface 
brightness, but since we prefer to work with individual counts and not smooth 
the surface brightness in any way, we replaced all the integrals with 
appropriately weighted sums over counts.

The powers are given by:
\begin{equation}
P_0 = \left[ a_0 \ln (R_{ap}) \right]^2
\end{equation}
\begin{equation}
P_m = \frac{1}{2 m^2 R_{ap}^{2m}} \left( a_m^2 + b_m^2 \right),
\end{equation}
where $R_{ap}$ is the aperture radius.
 The moments $a_m$ and $b_m$ are calculated using
\begin{equation}
a_m(R) = \sum_{r_i \le R_{ap}} w_i r_i^m \cos (m\phi_i)
\end{equation}
and
\begin{equation}
b_m(R) = \sum_{r_i \le R_{ap}} w_i r_i^m \sin (m\phi_i),
\end{equation}
where $r_i, \phi_i$  are the coordinates of the detected photon in polar 
coordinates and $w_i$ is its ``weight'' which is inversely proportional to the 
effective exposure at the given CCD location. The center of that polar 
coordinate system is chosen to set $P_1$ to zero.

In order to render the morphological information insensitive to overall X-ray 
flux, each of the angular moments $P_m, m=1,2,3...$ is normalized by the value 
of $P_0$, forming the power ratios, $P_m/P_0$. The power ratios $P_2/P_0$, 
$P_3/P_0$, $P_4/P_0$ have been used to characterize cluster substructure 
\citep{Jeltema2005}. $P_3/P_0$ has been found to be the best characterization of 
``disturbance''.

Aperture choice is very important for power ratios as they are most sensitive to 
the substructure at the maximum radius. Values of 1 Mpc, 0.5 Mpc, $R_{500}$ have 
been used as aperture radii. We use $R_{500}$ as it allows more consistent 
comparison of clusters of different mass than a fixed physical scale as 
$R_{500}$ is a natural scale for clusters of all masses and redshifts. The other 
substructure statistics are also based on an $R_{500}$ aperture, therefore, our 
comparison of various substructure statistics is consistent.

As many authors have noted, the power ratios calculated by the formulas above 
give values for $P_m$ biased high due to photon noise.
 This can be easily seen in the case of a perfectly symmetrical cluster - the 
random distribution of the angles $\phi_i$, and nonnegativity of $P_m$, lead to 
a distribution of $P_m$ with nonzero mean.
 Different authors used different methods to account for these biases.
We based our method of bias correction on the work of \citet{Bohringer2010a}, 
where the bias was computed by randomizing the polar angles for all collected 
photons, but keeping their radial distance fixed. The mean of the power $P_m$ of 
the mock observations obtained this way is interpreted as the typical photon 
noise contribution to the measurements of $P_m$ and subtracted from the $P_m$ of 
the real observation. We did not perform Monte-Carlo simulations for randomizing 
polar angles, because the mean of $P_m$ with randomized angles (uniformly 
distributed $\phi_i$) can be easily calculated analytically:
\begin{equation}
\begin{gathered}
n_m = \langle a_m^2 \rangle = \left\langle \left[ \sum w_i r_i^m \cos (m\phi_i) 
 \right]^2 \right\rangle \\
 =  \sum w_i^2 r_i^{2m} \langle \cos^2 (m\phi_i) \rangle = \frac12  \sum w_i^2 
 r_i^{2m} \\
\end{gathered}
\end{equation}
We need to subtract this value from both $a_m^2$ and $b_m^2$ which results in 
 the following formula for $P_m$.
\begin{equation}
P_m = \frac{1}{2 m^2 R_{ap}^{2m}} \left( a_m^2 + b_m^2 - 2n_m \right),
\label{eq:power_noise_corr}
\end{equation}
 
 After bias correction the background counts do not contribute to the powers $m 
 \ne 0$, but still contribute to $a_0 = \sum w_i$.
 To make $P_0$ and, consequently the ratios background independent, we need to 
also subtract the background contribution from $a_0$:

\begin{equation}
    a_0 = \sum w_i - w_{\mathrm{bkgr}}(R_{ap}),
\end{equation}
where $w_{bkgr}(R_{ap})$ is the expected total weight of all background photons 
within the aperture $R_{ap}$.

\subsection{Centroid shifts}

Centroid shift is another popular measure of ``disturbance'' of clusters.
It is defined by the variance of ``centroids'' obtained  by minimization of 
$P_1$ within 10 apertures($r \le n \times 0.1\, R_{500}$, with n = 1,2..10). The 
value of centroid shifts is expressed in units of $R_{500}$ which makes it a 
dimensionless quantity. Centroid shifts are defined slightly differently by 
different authors \citep[See][]{Mohr1995, Poole2006, OHara2006, Bohringer2010a}.  
Here we used
\begin{equation}
w =  \left[ {1 \over N-1} \sum_i \left( x_i -  \langle x \rangle \right)^2 
\right]^{1/2} \times {1 \over R_{500} }, \end{equation}
where $x_i$ is the position of the centroid of a given aperture.

\subsection{Concentration}
Concentration parameter is defined as the ratio of the peak over the ambient 
surface brightness.
Concentration has been widely applied to X-ray images \citep{Kay2008, 
Santos2008, Santos2010, Cassano2010, Hallman2011, Semler2012} and proved useful 
in distinguishing cool-core (CC) from non-cool-core (NCC) clusters. 

We adopted the definition of concentration provided by \citet{Santos2008}:
\begin{equation}
    c_{SB}= \frac{\Flux(r<40 \kpc,\; 0.5 \keV < E < 5 \keV)}
                 {\Flux(r<400 \kpc, \; 0.5 \keV < E < 5 \keV)}
\end{equation}
\label{csb}

The radii 40 and 400 kpc were chosen to maximize the separation between CC and 
NCC clusters. We computed concentration around the brightness peak as defined in 
Section~\ref{sec:centroid}, the same center that we used for photon asymmetry. 
Complete details on the stability of the concentration parameter can be found in 
\citet{Santos2008}.

\section{Photon asymmetry}

In this section we will describe our proposed morphological
classifier, namely photon asymmetry.

\subsection{Optical asymmetry and the motivation for photon asymmetry}

 In optical astronomy the asymmetry parameter is a part of the ``CAS'' galaxy 
classification scheme which stands for concentration (C), asymmetry (A) and 
clumpiness (S) \citep{Conselice2003}. Asymmetry quantifies the degree to which 
the light of an object (galaxy) is rotationally symmetric.
 It is measured by subtracting the galaxy image $I_{180}$ rotated by 
$180^{\circ}$ from the original image $I_0$ \citep{Conselice2003}:
\begin{equation}
A = \frac{|I_0-I_{180}|}{I_0}
\label{eq:optical-asym}
\end{equation}

 This definition tests \emph{central} (or \emph{mirror}) asymmetry, i.e. whether 
the image is invariant under a ``point reflection'' transformation (which is 
equivalent to rotation by $180^{\circ}$ around the central point). Although this 
definition of asymmetry has been applied to X-ray images of clusters before 
\citep[e.g.][]{Rasia2013}, it is only reliable for observations where the number 
of counts in each (binned) pixel is $\gg 1$. This condition is not satisfied for 
most cluster observations.

 One can come up with a similar definition of \emph{circular} or \emph{axial} 
asymmetry which would test whether the image is invariant under
rotation by arbitrary angle around the central point.
 That would involve finding the average intensity of the image in concentric 
annuli, and comparing local intensity with the average intensity in the annulus.
\begin{equation}
A = \int_0^{R} rdr \int d \phi \left(I(r,\phi) - \overline{I(r)}\right)^2
\label{eq:central-asym}
\end{equation}

This could also be a good measure of substructure and indeed people have tried 
to apply similar ideas for substructure statistics 
\citep[e.g.][]{Andrade-Santos2012}.

 The above definitions of asymmetry, both \eqref{eq:optical-asym} and 
\eqref{eq:central-asym}, are hard to implement for distant clusters whose 
observations have fewer counts.
We could generate smoothed images of clusters and apply the above definitions to 
these images, but that can generate biases. The large radial variations in 
surface brightness and the presence of substructure deny the ability to choose a 
single, global optimal smoothing scale. We cannot use an adaptive scale either, 
because asymmetry is then strongly dependent on the details of the adaptive 
smoothing algorithm.
 Also, by producing smoothed images (with either a fixed or adaptive scale), we 
effectively introduce some model-dependent prior on cluster structure. We would 
prefer, however, to only use objective information: the positions (and possibly 
energies) of the detected photons.

 Fortunately, there is a way to adapt the definition of asymmetry so that it can 
be computed efficiently in the limit of low photon counts, which we present in 
this paper. This adaptation is possible for both \emph{central} and \emph{axial} 
asymmetry. Central asymmetry might seem preferable, because it would have a zero 
value for a relaxed, but elliptical cluster. However, in our sample with few 
counts and ill-defined ellipticities, the values of axial and central 
asymmetries correlate strongly. Additionally, \emph{axial} asymmetry is 
conceptually simpler for our statistical framework, so we concentrate on it for 
this paper.

 Our strategy for adapting Eq.~\eqref{eq:central-asym} to the case of few counts 
with known coordinates is the following.
We split the image into a few annuli, and check whether the surface brightness 
is uniform in each of these annuli.
 In the limit of few counts, this is the same as checking whether these counts 
are uniformly distributed in the annulus. This amounts to  checking that their 
polar angles are uniformly distributed in the $0 \le \phi < 2\pi$ range.

\subsection{Photon asymmetry within an annulus}

\begin{figure*}[htbp]
    \epsscale{1}
    \plotone{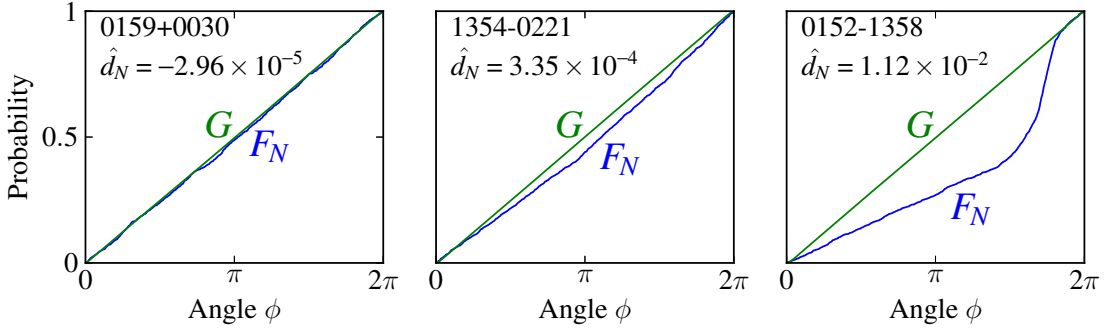}
    \caption{
The method described in this paper compares the observed cumulative probability 
distributions of the angular positions of photons (F$_N$), to that of a uniform 
distribution (G). This Figure shows empirical $F_N$ and uniform $G$ distribution 
functions in the outermost annulus for 3 progressively more disturbed clusters 
(0159+0030, 1354-0221, and 0152-1358). The more disturbed clusters manifest 
greater differences between $F_N$ and $G$ and correspondingly higher values of 
distance metric $d_N$.
}
    \label{fig:watson_examples}
\end{figure*}

 To assess the degree of nonuniformity of the angular distribution of the 
counts, we use Watson's test \citep{Watson61}. Watson's test compares 2 
cumulative distribution functions. Other members of this family of 
non-parametric tests for the equality of distribution functions include the 
well-known Kolmogorov-Smirnov test as well as less well-known Cramer-von Mises 
and Kuiper's tests. For the reasons explained in the Appendix, Watson's test is 
the only one that works in our specific situation. Unfortunately, Watson's test 
is only able to test the null hypothesis, i.e. compute the probability that the 
given sample is drawn from the assumed distribution.
 Our case is slightly different - we know that our sample (of counts as function 
of polar angles) is not drawn from the uniform distribution, so, in principle, 
goodness of fit tests are not applicable to our case.  However, as we show in 
the Appendix, we can interpret the value of Watson's test as the estimate of the 
distance between the true underlying distribution function and the assumed 
distribution function.  

Let us consider the photons that arrive in an annulus $R_{in} < r < R_{out}$ 
relative to the cluster center. The specific definition of these annuli will be 
discussed in Section~\ref{sec:choose-annuli}. Let $\Phi$ be a polar angle 
(random variable) of a cluster photon in the chosen coordinate system, centered 
on the cluster, and $\phi_1, \phi_2, \cdots, \phi_N$ are the polar angles of the 
observed photons in the annulus ($N$ = total number of observed photons in the 
annulus).  Then we will define:
\begin{equation}
F(\phi) = \mathrm{Prob}(\Phi \le \phi)
\end{equation}
as the true angular (cumulative) distribution function and \begin{equation}
\begin{gathered}
    F_N(\phi) = \frac{1}{N} \sum \left\{ \begin{array}{ll}
            1, & \textrm{ if } \phi_i \le \phi \\
            0, & \textrm{ otherwise}
                                     \end{array}
                             \right.        \end{gathered}
\end{equation}
as the measured (empirical) distribution function. Being distribution functions 
on a circle, $F$ and $F_N$ also depend on the arbitrary starting point $\phi_0$ 
which we write as
\begin{equation}
\begin{gathered}
    F = F(\phi; \phi_0) \\
    F_N = F_N(\phi; \phi_0).
\end{gathered}
\end{equation}

We can now introduce Watson's statistic $U_N^2[F_N,F]$ as
\begin{equation}
\begin{gathered}
    U_N^2[F_N, F; \phi_0] = N \int \Big(F_N(\phi; \phi_0) - F(\phi; 
    \phi_0)\Big)^2 dF \\
    U_N^2[F_N, F] = \min_{\phi_0} U_N^2[F_N, F; \phi_0],
\end{gathered}
\end{equation}
i.e. $U_N$ is the minimum value of integrated squared difference between $F_N$ 
and $F$ over possible starting points $\phi_0$.

The greater the value of $U_N^2$ the less likely that $F_N$ is produced by 
drawing from $F$. In our case $F$ is unknown, but we can test how likely it is 
that $F_N$ is drawn from another distribution $G$ which represents an idealized 
axisymmetric source. ($G$ would be uniform in the absence of instrumental 
imperfections)
\begin{equation}
    U_N^2[F_N,G] = N \min_{\textrm{origin on the circle}} \int(F_N-G)^2dG.
\end{equation}
Interestingly, it is possible to interpret $U_N^2[F_N, G] / N$ as the distance 
between $F$ and $G$:
\begin{equation}
    U_N^2[F_N,G] / N = \textrm{distance}(F,G) + \frac{1}{12N} +\textrm{Noise},
\label{eq:U-distance}
\end{equation}
where the bizarre term $1/12N$ comes from the properties of the statistic 
distribution under the null hypothesis. The detailed derivation of 
\eqref{eq:U-distance} is presented in the Appendix. Here we would like to note 
that the mean value of $\textrm{Noise}$ is smaller than $1/12N$ for the relevant 
values of $N$, therefore 
\begin{equation}
    \hat{d}_N = U_N^2 / N - 1 / 12N
\end{equation}
is an estimator of $\textrm{distance}(F,G)$, the distance between the observed 
and uniform distributions of photons in the annulus. The variance of this 
estimator scales as $1/N$, so that we can get better estimate of the distance as 
N increases.

The method is illustrated in Fig.~\ref{fig:watson_examples} where we show $F_N$, 
$G$, and the value of $\hat{d}_N$ in the outermost annulus for 3 progressively 
more disturbed clusters. The more disturbed clusters manifest greater 
differences between $F_N$ and $G$ and, consequently, higher values of 
$\hat{d}_N$.

As we are interested in the distance between the observed and a uniform 
distribution of \emph{cluster-only} photons (as opposed to \emph{cluster and 
background} photons), we additionally need to multiply that distance by the 
squared ratio of total counts $N$ to cluster counts $C$ in that annulus. As the
number of cluster counts $C$ is not directly observable, we estimate it by 
subtracting the expected number of background counts in the annulus from the 
total counts $N$. The resulting background-corrected expression   
\begin{equation}
    \hat{d}_{N,C} = \frac{N}{C^2} \bigg(U_N^2 - \frac{1}{12}\bigg),
\label{eq:d_def}
\end{equation}
is an estimate of the distance between the true photon distribution and the 
uniform distribution. (see Appendix for details)

\subsection{How to choose the optimal annuli}
\label{sec:choose-annuli}

    The first step in choosing optimal annuli is to select the maximum aperture 
radius. $R_{500}$ is a good choice, because cluster X-ray emission is typically 
indistinguishable from the background beyond this radius.  Also, we exclude the 
region $r < 0.05R_{500}$ from the analysis because pixelation artefacts at small 
radii distort Watson's statistic.

Second, we need to choose the number of annuli inside this $0.05R_{500} < r < 
R_{500}$ region.  One can use any number of annuli for the computation of 
asymmetry.  The tradeoff is between the asymmetry S/N in each individual annulus 
and radial resolution.  We found that it is desirable to have at least a few 
hundred counts in each annulus, so we used 4 annuli for our sample of clusters.  
This optimization may be different for a cluster sample with different number of 
counts.

 Finally, we need to choose the radii of these annuli. The relative uncertainty 
of asymmetry is estimated to be $\sqrt{N}/C$, where $N$ is the total number of 
counts, and $C$ is the number of cluster counts. The radial binning should be 
chosen carefully, because low numbers of $N$ or $C$ in any bin inflate the
uncertainty. This is a nontrivial task as the radial brightness profile is very 
different between clusters. We choose the radial binning to achieve a uniform 
relative uncertainty in asymmetry, for each annulus, across the clusters in our 
training set. The following choice of boundaries in units of $R_{500}$: 0.05, 
0.12, 0.2, 0.30, 1, leads to the most uniform uncertainty across bins. We 
caution that applying this technique to X-ray survey instruments or data sets 
that exhibit a broader PSF than Chandra's 0.5 arcsec FWHM will require a careful 
reevaluation of radial binning, sence we desire annuli widths $dr \gg 
\rm{FWHM}$.

The last step in the computation of photon asymmetry is to combine the values of 
 asymmetry from the 4 annuli. We use a weighted sum of distances  from each 
 annulus $\hat{d}_{N_k,C_k}$
(see Eq.~\eqref{eq:d_def}, $k$ numbers the annuli, $N_k$ and $C_k$ are the total 
and cluster counts in $k$-th annulus) with a weight equal to the estimated 
number of cluster counts $C_k$ in that annulus.  \begin{equation}
    A_{phot} = 100 \sum_{k=1}^4 C_k \hat{d}_{N_k, C_k} / \sum_{k=1}^4 C_k
\end{equation}
 
We introduced a multiple of 100 into the definition of $\rm{A_{phot}}$ to bring 
all the asymmetries to a convenient range $0 < \rm{A_{phot}} \lesssim 3$. The 
resulting quantity is independent of exposure and background level.

\subsection{Cluster centroid determination}
\label{sec:centroid}
 The standard prescription for optical asymmetry is to choose the center that 
minimizes asymmetry. However this method is prone to producing values of 
asymmetry that are biased low. This effect is especially noticeable in our 
resamplings with very low number of counts.

We based our choice of centroiding on three considerations: 1) we favor a 
centroid choice that is independent of the asymmetry computation, 2) if the 
cluster possesses a strong core, we use that feature to define the cluster 
center, and 3) by assigning the cluster center to a high S/N region of the 
image, we can compute asymmetry in annuli at high S/N.

Based on these requirements, we chose the center to be the brightest pixel after 
 convolution with a Gaussian kernel with $\sigma = 40$ kpc.
 At $z=1$, a single Chandra pixel corresponds to about 4 kpc, and the Chandra 
PSF FWHM is of order 2 pixels, so the smoothing scale is much coarser than 
Chandra angular resolution.

The centroid defined as a convolution with a Gaussian kernel is not very 
sensitive to the size of this Gaussian kernel. We chose the kernel size to be 
40kpc to be consistent with our definition of concentration. We use this 
centroid for both asymmetry and concentration. We stress that the 
Gaussian-convolved image is used only for centroiding, not for computation of 
any substructure statistics.

\subsection{Additional remarks}
The $A_{phot}$ parameter can be applied beyond X-ray observations. In fact, a 
quantity defined exactly the same way can be used for optical observations of 
clusters if we replace the coordinates of each photon with the coordinates of a 
member galaxy. Additionally, each galaxy can be given a weight that depends on 
its optical luminosity (e.g. w $\sim \log$ luminosity).  These weights can 
replace the ones in eq.(13), so that F remains a proper distribution function.  
In this case the equations (13)-(20) remain valid yielding a parameter that 
describes the asymmetry of the galaxy distribution within a cluster derived from 
optical observations.

\bigskip
\section{Simulated observations and determination of uncertainties}
\label{sec:uncertainties}

\subsection{Simulated observations}
We now address the questions of 1) sensitivities of substructure statistics to 
observation parameters, and 2) uncertainties of these substructure statistics, by 
calculating them for simulated observations with the desired parameters (such as 
exposure or background level). The idea of using simulated observations in 
similar ways goes back to the works of \citet{Buote&Tsai1996, Jeltema2005, 
Hart2008, Bohringer2010a}. Generating these simulated observations is 
straightforward if we have the map of the true cluster surface brightness (or, 
more precise, cluster brightness multiplied by CCD exposure map) - we would draw 
each pixel value from the Poisson distribution with the mean equal to that 
brightness. As we don't know that true underlying brightness distribution, we 
use instead our best approximation to it, which is the result of an adaptive 
smoothing algorithm. 

To simulate changing the exposure, before drawing from Poisson distribution, we 
need to multiply the surface brightness map by a constant; to change the level 
of background, we need to add a constant to the surface brightness map; to 
change the telescope PSF, we need to convolve the existing brightness map with 
the new PSF (the real Chandra PSF is negligibly small).

To simulate how the clusters would look if they were moved to a greater 
redshift, we need to calculate the expected X-ray flux from that cluster, 
rescale the number of observed counts accordingly, change the image spatial 
scale (which is a small correction as angular diameter distance doesn't change 
much from z = 0.3 to 1), and then increase the amount of the background to its 
old value.
The only tricky part in this process is the calculation of the new cluster flux 
which should include the change in the luminosity distance, and the K-correction 
\citep{Hogg2002} that compensates for the shift in the cluster emission in the 
observed frame.
\begin{equation}
\frac{\Flux_{\new}}{\Flux_{\old}} = \frac{D_{L,\old}^2}{D_{L,\new}^2} 
\frac{K(z_{\new})}{K(z_{\old})}
\end{equation}
Since we don't need to simulate this very precisely -- we only want to get an 
idea of how it affects the substructure measures -- we use a simple 
approximation to \citet{Santos2008} results for 0.5-5 keV energy band:
\begin{equation}
    K(z) = \frac{1}{1+2z}
\end{equation}

\subsection{Uncertainties}
\label{sec:uncertainties}

To estimate the uncertainties of the various substructure statistics, we used the 
above-described algorithm to generate 100 mock observations with exactly the same 
exposure and background level as in the original observations, but varied noise 
realization. Then we computed the substructure statistics for these samples, and 
found the median, the 16th lowest and the 16th highest observed value in the 
sample. We treat the median as the characteristic central value of statistic for 
this set of mock observations, and the interval between 16th lowest and the 16th 
highest observed value as the 1$\sigma$, or 68\% confidence interval. Using 
order statistics for the central value and the confidence interval is the most 
sensible choice for us, because the distributions for any substructure statistic 
values are asymmetric and extremely heavy tailed.  The statistic value obtained 
from the real observation didn't always fall within this confidence interval for 
two reasons.  First, as this is only a 68\% confidence interval, we expect 
approximately $1/3$ of all points to be outside of the $1\sigma$ range.  Second, 
the resampling process tends to overestimate the cluster substructure.  This 
arises because our smoothed surface brightness maps do contain some residual 
noise due to Poisson statistics from the cluster and the background, and we then 
inject an additional component of shot noise when computing a fake cluster 
observations.   Thus, the value of the statistic for mock observations may be 
biased, and the confidence intervals for the mock observations and the real ones 
are not expected to coincide.  However, we expect that the true surface 
brightness and the inferred one would produce the samples of statistic values 
with similar variances.  Therefore, we can use the variability of the simulated 
sample to determine the size of the error bars, but should center the error bars 
on the statistic value obtained for the real observation instead of the mean of 
the sample. A similar method of calculating uncertainties from simulated 
observations was used by \citet{Bohringer2010a}.

\begin{figure}[htbp]
    \epsscale{1}
    \plotone{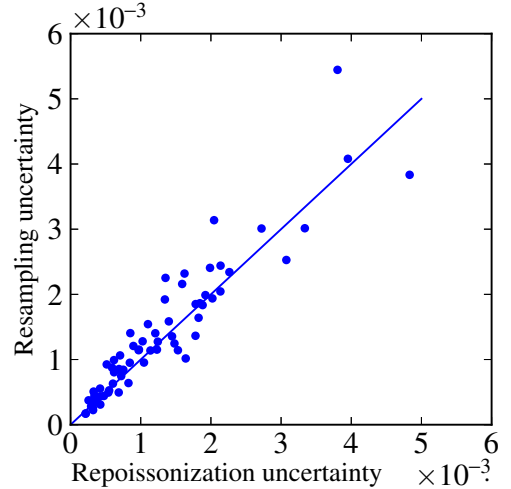}
    \caption{
Comparison of A$_{\rm{phot}}$ uncertainties computed by two different methods.  
The horizontal axis represents uncertainties estimated by ``repoissonization'' 
\citep{Weissmann2013}.  The vertical axis represents uncertainties estimated by 
resampling half of the observation photons with replacement. The two methods 
agree well, suggesting that the simpler of the two (resampling) is a sufficient 
representation of the ``true'' uncertainty.
}
\label{fig:errbar-cmp}
\end{figure}

    The method described above provides robust uncertainty estimates, but 
requires   complicated machinery that generates adaptively smoothed maps and 
mock observations. We have used this machinery to perform substructure 
sensitivity tests, but one may want to use simpler uncertainty estimation 
methods when only interested in the uncertainty of asymmetry for a given 
observation. Therefore, we developed a simplified uncertainty estimation method 
which does not use the adaptive smoothing algorithm. We used subsampling method 
to determine the scatter in the measured asymmetry values. We generated mock 
observations that take a random half of the counts from the original observation 
and computed substructure statistics from them. The scatter in the resulting 
asymmetry values is expected to $\sqrt{2}$ larger than what we would obtained 
for the full sample, so we need to reduce these error bars by $\sqrt2$. This 
method avoids additional assumptions about clusters introduced by the adaptive 
smoothing algorithm, and is significantly simpler in implementation.  We 
compared the error bars produced with both methods (Fig.  \ref{fig:errbar-cmp}), 
and found them to be similar.

We also produced samples of 100 mock observations each where we changed one 
parameter of observation (such as exposure) for our sensitivity tests. In these 
tests, we viewed the adaptively smoothed images of clusters as the true surface 
brightness distributions in the sky. Unlike the previous group of simulations, 
here the true value of the statistic is not relevant. The median and the 68\% 
confidence interval for each such sample represent how the statistic reacts to 
the corresponding change in the parameter of observation (such as exposure).

\begin{turnpage}
\begin{figure*}
    \epsscale{1.05}
    \plotone{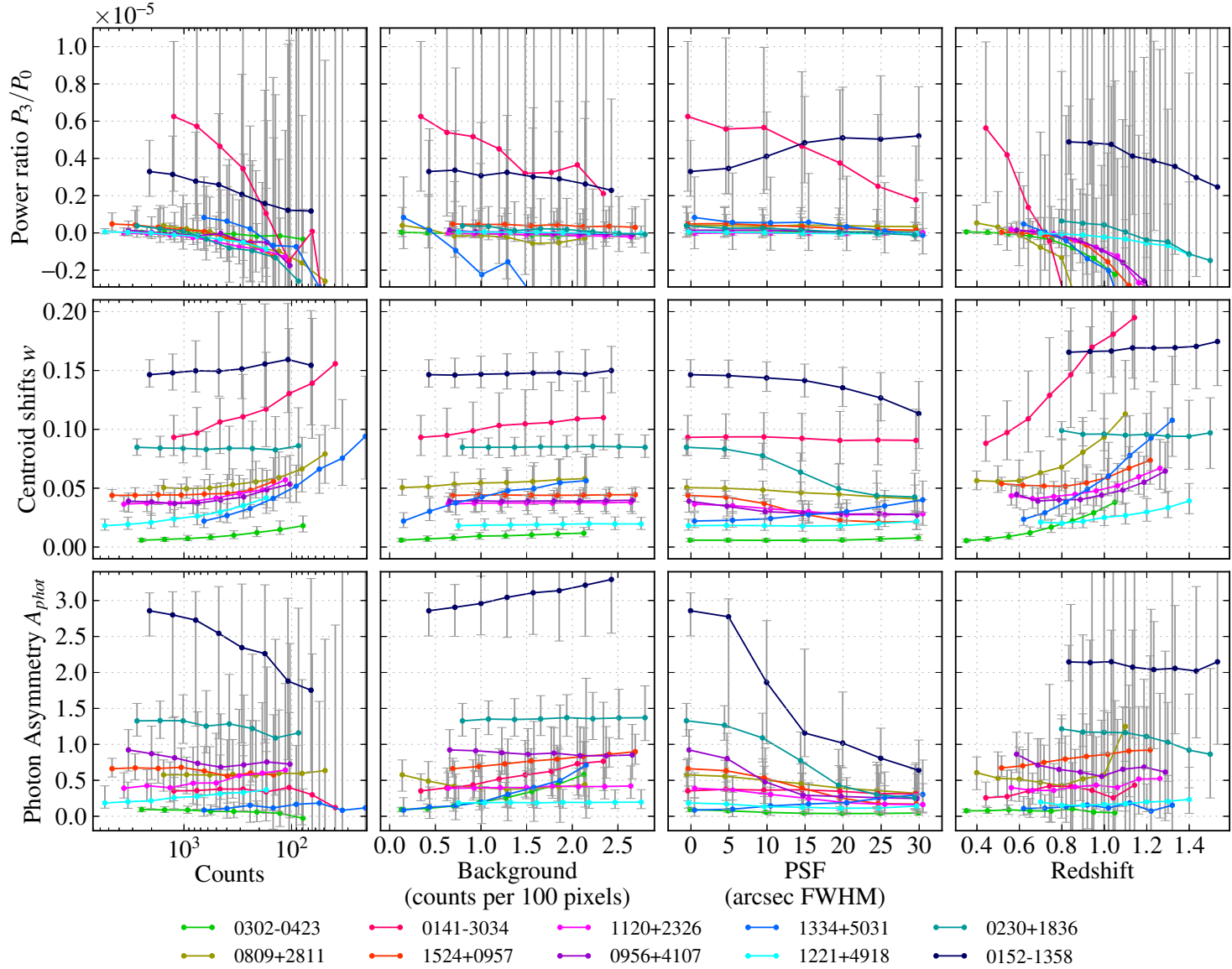}
    \caption{
        Sensitivity of 3 substructure statistics: power ratios, $P_3/P_0$ (1st 
        row), centroid shifts, $w$ (2nd row), and photon asymmetry, 
        $\rm{A_{phot}}$ (3rd row) to the quality of observations: number of 
        counts within $R_{500}$ (1st column),
        background level (2nd column), telescope PSF (3rd column), and
        cluster redshift at fixed exposure (4th column).
        Only a representative subset of the clusters is shown in this Figure - 
        the complete sample is shown in Fig.\ \ref{fig:sens-plots-all}. An 
        idealized morphological statistic would be insensitive to the quality of 
        observations, i.e.\ all the lines should be parallel to x-axis.
        The photon asymmetry parameter presented in this paper shows 
        better stability and better resolving power for observations of poor 
        quality than commonly-used power ratios and centroid shifts.
        The names of the clusters are identical to those used in \citet{CCCPII}.
    }
    \label{fig:sens-plots-sel}
\end{figure*}

\begin{figure*}
    \epsscale{1.05}
    \plotone{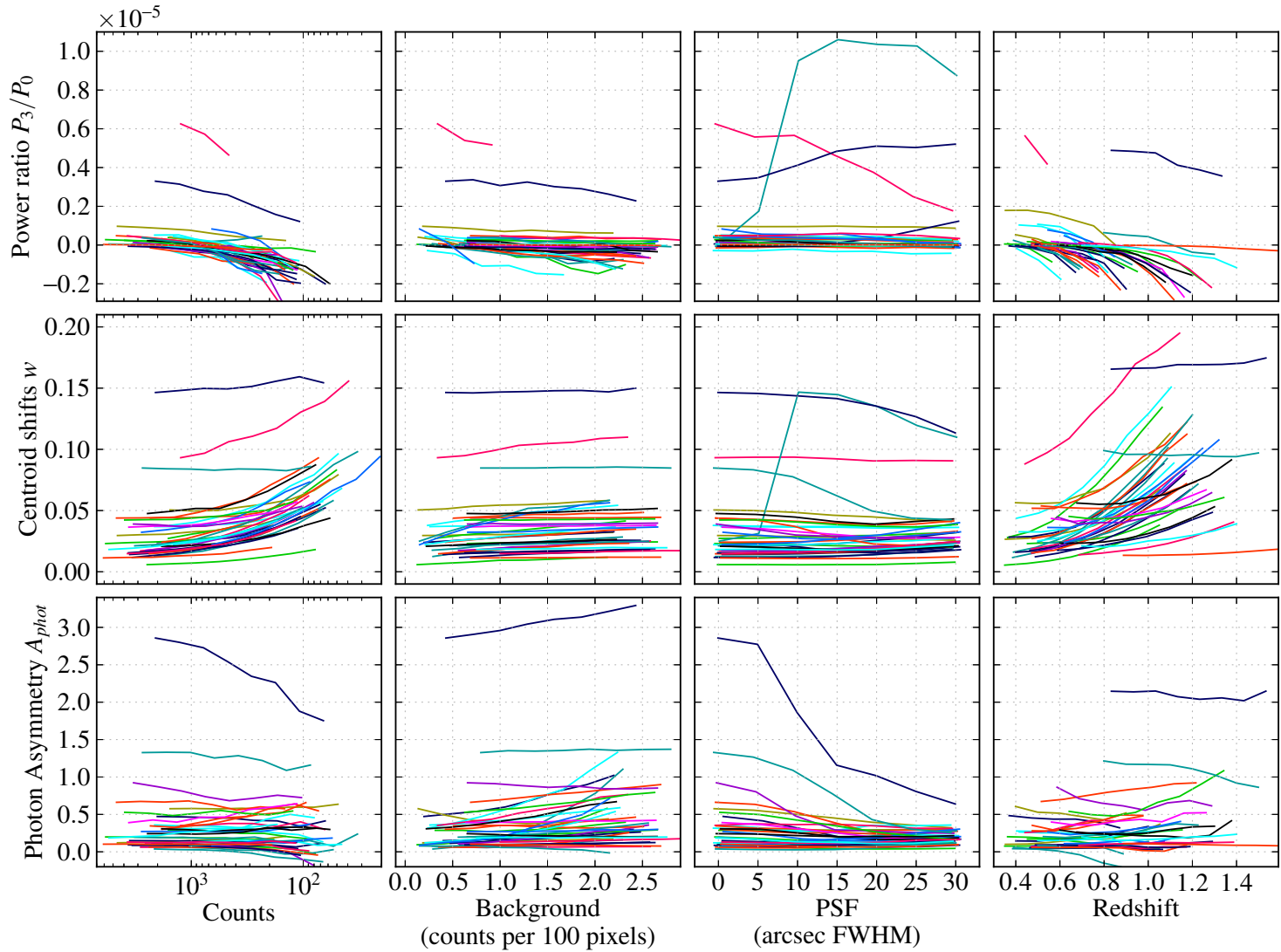}
    \caption{
        Similar to Fig.~\ref{fig:sens-plots-sel}, but now showing the full 
    sample of 36 clusters from \citet{CCCPII}. Uncertainties are excluded in 
    this plot in order to reduce clutter -- see Fig.~\ref{fig:sens-plots-sel} 
    for a subset of these clusters with uncertainties. The points where 
    uncertainties exceed the entire dynamic range for the corresponding 
    substructure statistic are excluded from the plots.
    }
    \label{fig:sens-plots-all}
\end{figure*}
\end{turnpage}

\begin{figure*}[htbp]
    \epsscale{1.2}
    \plotone{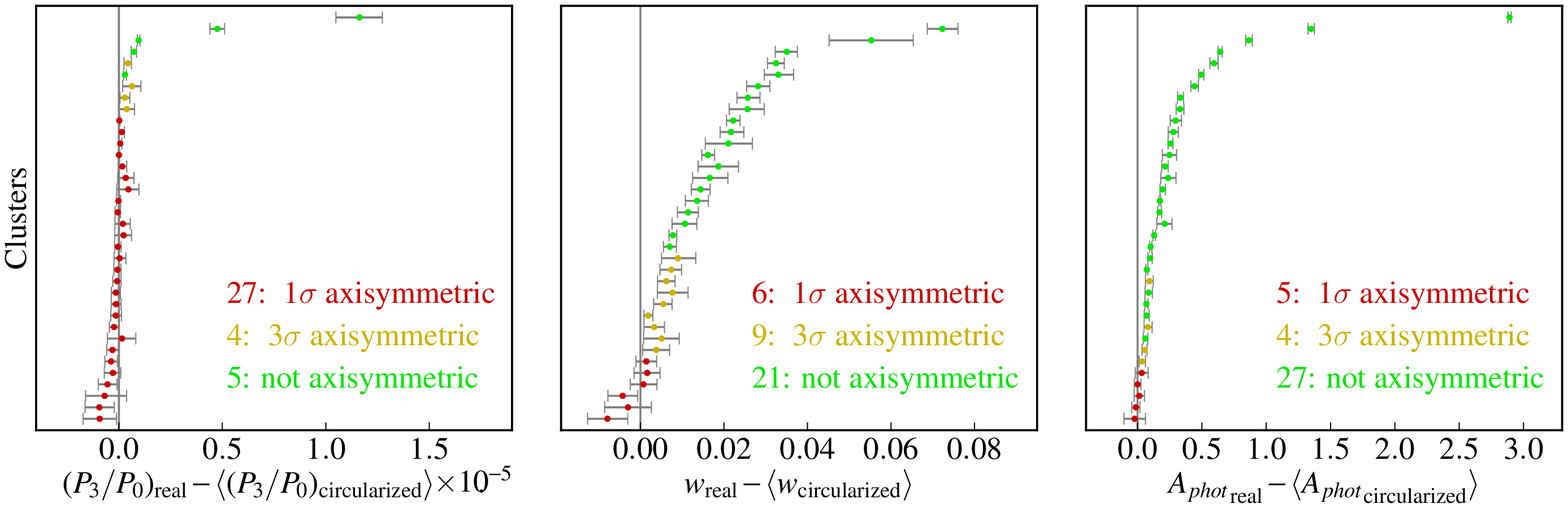}
    \caption{
This figure demonstrates how well each of the three substructure statistics 
examined in this paper is able to discriminate between the true photon 
distribution and the idealized case of axisymmetry for each of the 36 clusters 
in \cite{CCCPII}. For each cluster, we compare the measured substructure 
statistic to the expectation for a perfectly smooth, axisymmetric source.  
Horizontal bars represent 1$\sigma$ confidence intervals. The red points 
indicate clusters that are within $1\sigma$ of axisymmetry and, thus, are 
indistinguishable from the unrealistic case of perfect axisymmetry. The yellow 
and green points indicate clusters that are within $3\sigma$ and $>$3$\sigma$ of 
axisymmetry, respectively. Since no cluster should be \emph{perfectly} 
symmetric, this plot demonstrates the sensitivity of each statistic to low 
levels of substructure, with A$_{\rm{phot}}$ performing the best overall.
}
    \label{fig:circularized}
\end{figure*}

\bigskip
\section{Results and Discussion}

\subsection{Sensitivity of morphological parameters to data quality}
\label{sec:SNRsensitivity}

An important test for any substructure statistic is its insensitivity to the 
observational S/N. Here we present sensitivity tests of two currently-popular 
substructure parameters (centroid shifts and power ratios $P_3/P_0$) and the new 
one introduced in this paper (photon asymmetry, $\rm{A_{phot}}$). We conducted 4 
tests that degraded the observations in different ways, namely 1) reduced the 
number of photons (exposure), 2) increased the level of background, 3) 
``blurred'' the observation with larger PSF (or, alternatively, decreased the
cluster's angular size), and 4) altered the observations in all mentioned ways, 
simulating an observation of the same cluster with the same exposure as if it 
was at higher redshift.
 
The plots of all sensitivities are presented in 
Figures~\ref{fig:sens-plots-sel} and \ref{fig:sens-plots-all},
with different statistics in rows and sensitivity tests in columns. 
Fig.~\ref{fig:sens-plots-sel} shows both median values of statistics and 
$1\sigma$ confidence intervals, but only for a subset of representative 
clusters, while Fig.~\ref{fig:sens-plots-all} shows only median values of 
statistics that we obtained in our Monte-Carlo simulations (see 
Sec.~\ref{sec:uncertainties}).  We chose to present plots of only $P_3/P_0$ for 
power ratios, because $P_3/P_0$ is believed to be the best indicator of 
substructure. The plots of $P_2/P_0$ and $P_4/P_0$ look qualitatively very 
similar.

All statistics show relative insensitivity to the number of cluster counts, at 
levels above $\sim$ 2000 counts. However, in the low counts regime, both power 
ratios and centroid shifts show strong biases. The power ratio tends to be 
biased low for all clusters. Centroid shifts tend to be biased high, more so for 
clusters that do not show significant substructure. Each cluster seems to have 
its own threshold value in number of counts, so that centroid shifts are stable 
when there are sufficient cluster counts, but start to increase as the simulated 
number of counts falls below this threshold value. This behavior of centroid 
shifts is not surprising, because the statistical error of finding a centroid of 
a few photons should scale as one over the square root of the number of photons, 
unless there are significant secondary emission peaks that ``pin'' centroids of 
certain radii. In other words, although this bias has a similar behavior for 
many clusters, it cannot be corrected simply as a function of number of counts 
-- it also depends on the morphology \citep{Weissmann2013}.  

Centroid shifts, perhaps unsurprisingly, are the most stable statistic with 
respect to background levels. The determination of centroid is simply 
insensitive to a uniform background (unless there are so few counts that the
$1/\sqrt{N}$ effect described in the previous paragraph starts playing a role).  
Power ratios are relatively stable with respect to background levels.  (Although 
they become consistent with zero for \emph{every} cluster in the sample after 
even a moderate background increment due to increased uncertainty.) Asymmetry is 
insensitive to background levels as long as a reliable estimate of cluster 
counts is possible in each annulus.  However, when the square root of the total 
counts becomes comparable to the cluster counts, the estimate of cluster counts 
may become close to zero (or even negative). This unphysical estimate of cluster 
counts, being in the denominator in Eq.~\eqref{eq:d_def}, drives the statistic 
to high absolute values. This is a drawback of $\rm{A_{phot}}$, which could be 
fixed by a more careful separation of background and cluster counts.  

None of the statistics are stable against PSF increase because at 
$30^{\prime\prime}$ PSF the substructure is completely washed out and 
undetectable by any method. Asymmetry has a stronger sensitivity to PSF, because 
it probes the non-uniformity of the photon distribution on all angular scales, 
starting from the lowest Fourier harmonics to the highest. Power ratio 
$P_3/P_0$, on the other hand, is only sensitive to the third Fourier harmonic.  
It is interesting that the PSF has a much stronger influence on any substructure 
statistic than does the number of counts. This observation suggests that for 
substructure studies a telescope's angular resolution is more important than its 
effective area. 

The redshift test is the most challenging: the luminosity distance increases 
very fast, and the K correction adds to the flux dimming, effectively making the 
high-z simulated observations dominated by the background.  Fluctuations in 
background increase variability of centroid estimation, driving centroid shifts 
to higher values. (A similar effect is demonstrated by sensitivity to cluster 
counts.) The power ratio median ``dives'' down to negative values (again, 
similar to counts test). Additionally, power ratio uncertainties increase very 
quickly, which is the result of background correction (subtraction of two nearly 
equal terms in  Eq.~\eqref{eq:power_noise_corr}). Photon asymmetry also suffers 
from background correction, but overall shows less sensitivity to simulated 
redshift than either power ratio or centroid shifts.

What sets photon asymmetry apart from power ratios and centroid shifts is much 
smaller relative uncertainties. Unlike power ratios and centroid shifts, photon 
asymmetry is typically further than one standard deviation away from zero. So, 
photon asymmetry is capable of separating relaxed and slightly unrelaxed cluster 
populations in the case of observations with even a few hundred X-ray counts. To 
demonstrate that photon asymmetry is better than its competitors at 
distinguishing the clusters that are inconsistent with axisymmetric sources in 
the low S/N regime, we calculated the number of clusters in our sample that are 
1$\sigma$ consistent with circularly symmetric sources.

In order to compare the statistical significance of three different
substructure parameters (photon asymmetry, centroid shifts and power ratios), 
for each cluster we generated a set of idealized, axisymmetric clusters. This was done 
by retaining the exact radial location for each of the N detected photons for 
each cluster, but with a random realization of polar angles for each photon's 
position.  We then computed the relevant substructure metric. For each cluster 
we subtracted the mean of the parameter values computed from the fake circular 
clusters from that obtained from the actual cluster. We also assigned the 
scatter in the fake measurements as the uncertainty, for each cluster.  
Figure~\ref{fig:circularized} shows departure from the circular case, with 
photon asymmetry clearly achieving a more significant determination of cluster 
substructure. Confidence intervals overlapping with 0 (red points) mean that the 
cluster indistinguishable from the axisymmetric case. The yellow points indicate 
clusters that are within $3\sigma$ of axisymmetry.

The number of clusters that are statistically inconsistent (at $3\sigma$) with 
the idealized, axisymmetric case, as determined by different substructure 
statistics, are as follows:
\begin{itemize}[itemsep=0pt]
  \renewcommand{\labelitemi}{$-$}
  \item Power ratio $P_3/P_0$: 5 (out of 36)
  \item Centroid shifts $w$: 21 (out of 36)
  \item Photon asymmetry $\rm{A_{phot}}$: 27 (out of 36)
\end{itemize}

In other words, photon asymmetry has the best resolving power to measure 
``disturbance'' in our sample.

\begin{figure}[htbp]
    \epsscale{1.2}
    \plotone{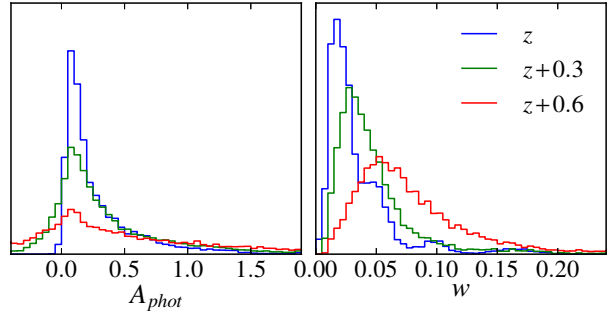}
    \caption{
        Distributions of substructure statistics for the entire sample of 
        simulated observations at different redshifts.
        (Left) The peak of $\rm{A_{phot}}$ distribution doesn't move, for 
        simulated observations at higher redshift.         
        (Right) The peak of $w$ distribution shifts to higher values as the 
        clusters are shifted to higher redshift.
        }
    \label{fig:distr_z}
\end{figure}

The tendency of centroid shifts to be biased high for low-counts observations 
makes it questionable whether it can provide any meaningful results for samples 
of clusters with nonuniform S/N. We tested how the properties of the entire 400d 
sample would change if every cluster were moved to a greater redshift. In   
Figure~\ref{fig:distr_z}, we plot the distributions of $w$ and $\rm{A_{phot}}$ 
for the entire set of simulated observations at the original redshifts (blue), 
at the redshift $z+0.3$ (green), and at the redshift $z+0.6$ (red). We can see 
from the Figure that although the scatter is greater for the cluster sample at a 
higher redshift, the peak of the $\rm{A_{phot}}$ distribution doesn't shift.  
This observation confirms that we can safely compare the values of asymmetry for 
cluster observations of significantly different S/N and redshifts. In 
Figure~\ref{fig:distr_z}, the situation is different for $w$: the peak in its 
distribution shifts significantly moving to higher redshift, creating the false 
impression that higher-redshift clusters are more disturbed than their 
lower-redshift counterparts.

  Overall, photon asymmetry is more stable with respect to changes in number of 
counts, background and redshift, and has smaller uncertainty than both centroid 
shifts and power ratios.  

\begin{figure*}[htbp]
    \epsscale{1.2}
    \plotone{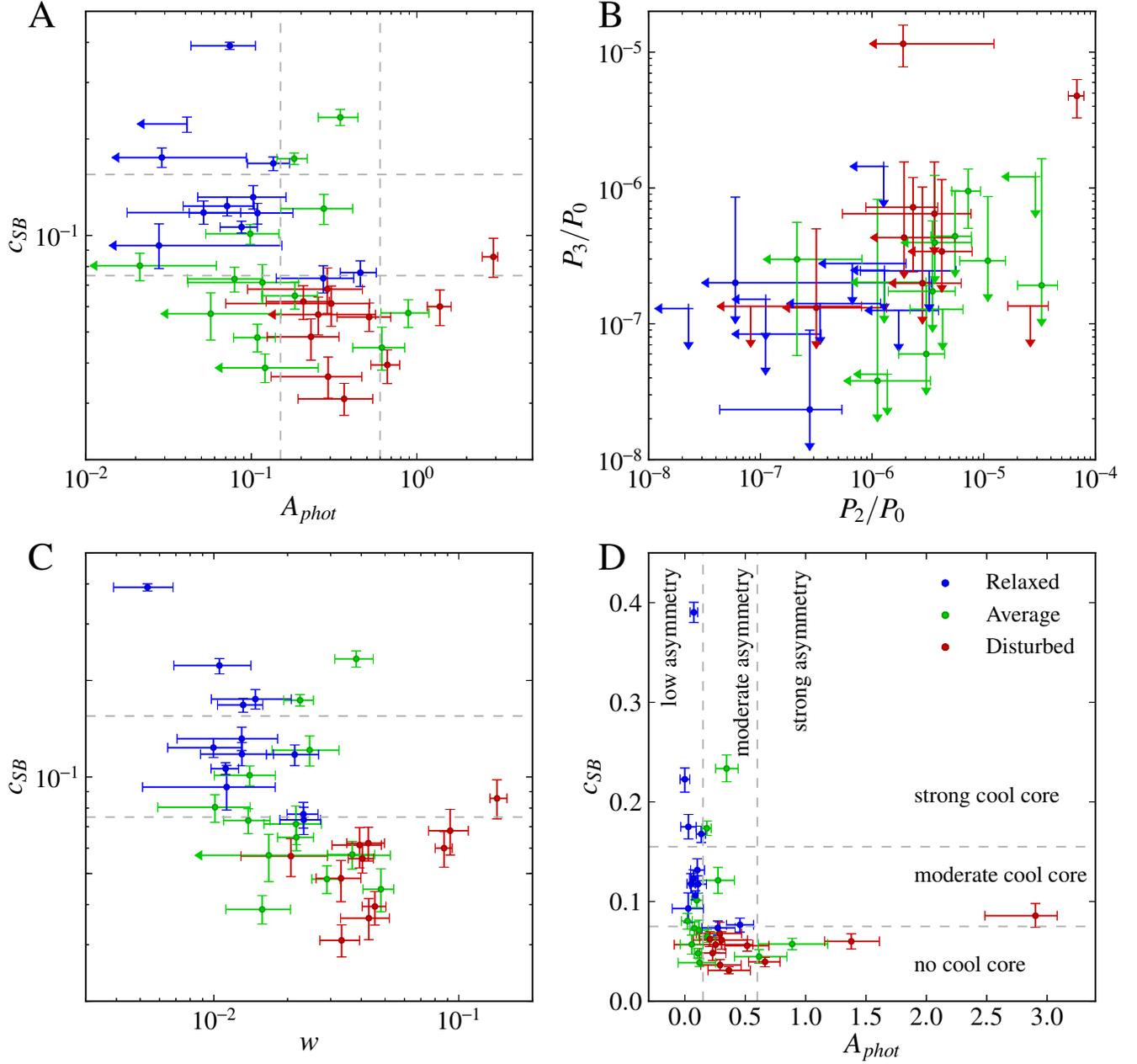}
    \caption{
        A: Cluster classification by A$_{\rm{phot}}$ (substructure statistic 
introduced in this paper) and surface brightness concentration 
\citep[c$_{\rm{SB}}$;][]{Santos2008}.  This classification scheme clearly 
separates relaxed, cool core clusters (high c$_{\rm{SB}}$, low A$_{\rm{phot}}$) 
from non-relaxed, disturbed systems (low c$_{\rm{SB}}$, high A$_{\rm{phot}}$).
        B: An alternative popular automatic classification scheme based on power 
ratios \citep{Jeltema2005} -- see text for details. Overall, the uncertainties 
are larger than in the asymmetry-concentration plane, and the clustering of 
same-type clusters is not as pronounced.  
        C: Cluster classification using centroid shifts  $w$ instead of 
$\rm{A_{phot}}$.  Panel C achieves similarly good separation of relaxed and 
disturbed clusters, however the values of $w$ may be correlated with 
observational S/N as shown in Section~\ref{sec:SNRsensitivity}.  
        D: Same as A, but in linear coordinates. This plot emphasizes that there 
are no clusters that are both highly concentrated and ``asymmetric'', consistent 
with the picture of cool cores representing a ``relaxed'' state that can be 
disrupted by cluster mergers.  
        The horizontal dashed lines splitting the range of concentrations 
correspond to cool core categories as defined by \citet{Santos2008}. The 
vertical dashed lines splitting the range of asymmetries correspond to asymmetry 
categories as defined here (low asymmetry: $\rm{A_{phot}}<0.15$; moderate 
asymmetry: $0.15 < \rm{A_{phot}} < 0.6$; strong asymmetry: $\rm{A_{phot}}>0.6$).  
In all panels, colors are assigned based on by-eye classification of 
``disturbance'' (see text for details): blue - most relaxed, green - average, 
red - most disturbed.  Where the corresponding values are consistent with zero 
they are plotted with arrows as their lower limits. Where they are negative, 
they are plotted with arrows at their upper limits. Values with negative upper 
limits are absent from the plots.  }
    \label{fig:classifications}
\end{figure*}


\subsection{Asymmetry-concentration diagram}

\begin{figure*}[htbp]
    \epsscale{1.2}
    \plotone{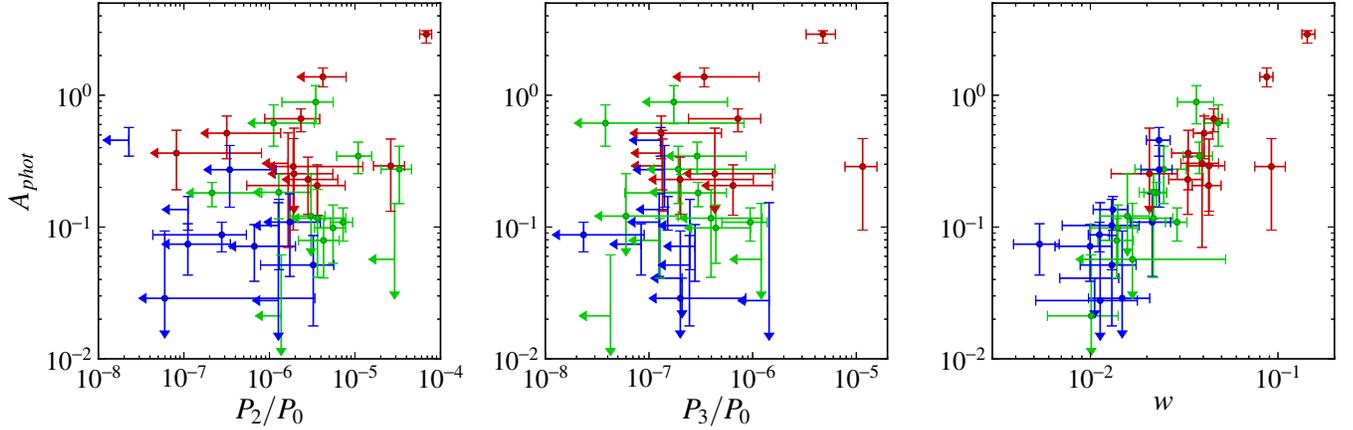}
    \caption{
Comparison of photon asymmetry, $\rm{A_{phot}}$, with power ratios, $P_2/P_0$ 
and $P_3/P_0$, and centroid shifts, $w$. Where power ratios are consistent with 
zero they are plotted with arrows as their lower limits.  Where power ratios are 
negative, they are plotted with arrows at their upper limits.  Power ratios with 
negative upper limits are absent from the plot. Colors are assigned based on 
by-eye classification of ``disturbance'': blue - most relaxed, green - average, 
red - most disturbed (see text for details). There is no obvious correlation 
between $\rm{A_{phot}}$ and the other substructure parameters, with the 
exception that they all tend to agree on the most disturbed systems.
    }
    \label{fig:asym_other}
\end{figure*}

\begin{figure*}[htbp]
    \epsscale{1.2}
    \plotone{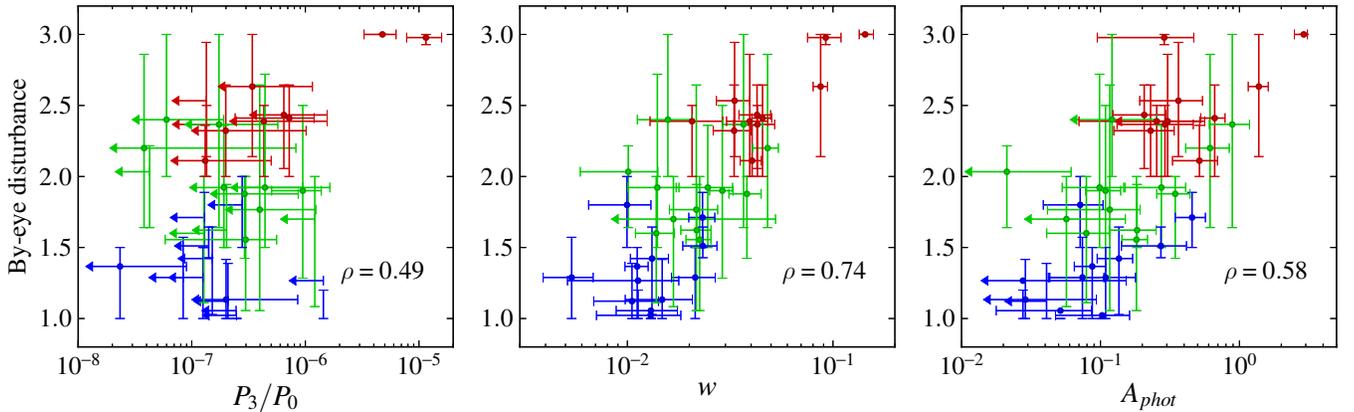}
    \caption{
        Comparison of 3 substructure statistics with the average by-eye 
        disturbance score as evaluated by a group of nine astronomers. The 
    correlation coefficient between the by-eye score and the corresponding 
statistic is shown on the plots. The colors based on by-eye disturbance score in 
the same way as in Fig.~\ref{fig:classifications} and \ref{fig:asym_other}.
}
    \label{fig:byeye_corr}
\end{figure*}

We propose a cluster classification scheme based on both concentration and 
asymmetry. Figure~\ref{fig:classifications}A shows the asymmetry-concentration 
diagram in logarithmic coordinates. The colors in Fig.\ref{fig:classifications} 
are based on cluster ``disturbance'' as evaluated by-eye by a group of nine 
astronomers. Each participant was asked to score the disturbance of the clusters 
on the scale 1 to 3 (fractional values allowed), with 1 being least disturbed 
and 3 being most disturbed. We found that 11 of the clusters were unanimously 
ranked in the most disturbed half. We call this group of clusters ``most 
disturbed'', and mark them in red in Figures~\ref{fig:classifications}, 
\ref{fig:asym_other}, and \ref{fig:byeye_corr}.  Another 12 clusters were 
unanimously placed in the least disturbed half of the rankings.  We call this 
group of clusters ``relaxed'' and mark them in blue. The remaining 13 clusters 
are ``average'' and marked in green.  

The asymmetry-concentration diagram (Fig.~\ref{fig:classifications}A) shows a 
significantly better separation of clusters at different states of dynamical 
equilibrium (as assessed by human experts) than the competing scheme of cluster 
classification based on power ratios proposed by \citet{Jeltema2005} and 
presented in Fig.~\ref{fig:classifications}B. The other drawbacks of the power 
ratios classification scheme are that $\log P_2/P_0$ and $\log P_3/P_0$ 
correlate (correlation coefficient = 0.61) and that both $P_2/P_0$ and $P_3/P_0$ 
are often consistent with 0. Therefore, what we see in 
Fig.~\ref{fig:classifications}B is mostly noise, whereas most clusters in
Fig.~\ref{fig:classifications}A show a significant detection of substructure as 
discussed above.

A similar separation of clusters at different states of dynamical equilibrium 
can be achieved using $w$ instead of $\rm{A_{phot}}$ as the substructure 
statistic (Fig.~\ref{fig:classifications}C), however $\rm{A_{phot}}$ is more 
stable and less biased for low S/N observations, as discussed above.

One can see that in Fig.~\ref{fig:classifications}A clusters avoid the upper 
right corner which confirms the standard assumption that concentrated or CC 
clusters are more regular. Fig.~\ref{fig:classifications}D, which is the same as 
Fig.~\ref{fig:classifications}A, but plotted in linear coordinates, shows this 
even better. It has a characteristic L-shape which implies that clusters are 
primarily either ``concentrated'' (upper part of the diagram) or ``asymmetric'' 
(right side) or ``normal'' (lower left corner).

In all the relevant panels of Fig.~\ref{fig:classifications} we plot two dashed 
vertical lines as threshold values that separate low-, medium-, and 
strong-asymmetry clusters.  The threshold values are $\rm{A_{phot}} > 0.15$ and 
$\rm{A_{phot}} > 0.6$. The horizontal dashed lines separate strong, moderate, 
and no cool cores as defined by \citet{Santos2008}. The threshold values are 
$\rm{c_{SB}} > 0.075$, and $\rm{c_{SB}} > 0.155$.

The asymmetry-concentration classification scheme makes a clear separation 
between the radial and the angular structure. Concentration only probes the 
radial photon distribution, while asymmetry probes the angular photon 
distribution. We expect these to be uncorrelated, a point to which the data 
attest (correlation coefficient = -0.20). We show how asymmetry compares with 
power ratios and centroid shifts in Fig.~\ref{fig:asym_other}. $\rm{A_{phot}}$ 
and $w$ are correlated strongly with correlation coefficient 0.87. This 
indicates that for high S/N data $\rm{A_{phot}}$ and $w$ agree well on which 
clusters are disturbed.

\subsection{Relative ranking of clusters by the amount of substructure, by-eye 
classification}

    In Fig.~\ref{fig:byeye_corr} we show how photon asymmetry, centroid shifts 
and $P_3/P_0$ power ratio compare to by-eye classification. We find that the 
photon asymmetry parameter correlates with the human ``by-eye'' ranks almost as 
strongly as centroid shifts, with a Spearman's rank correlation coefficient of 
0.71 for $\rm{A_{phot}}$, and 0.75 for $w$. Power ratio $P_3/P_0$, on the other 
hand, shows much lower correlation coefficient 0.47.

In Figures \ref{fig:gal-asym-w}, \ref{fig:gal-asym-P30}, 
\ref{fig:gal-asym-byeye}, we present three side-by-side comparisons of 
morphological indicators. In each figure the left panel shows the (same) X-ray 
images of galaxy clusters, ordered by increasing values of our photon asymmetry 
parameter. The right panel shows these same clusters, ranked by increasing 
centroid shifts, power ratio $P_2/P_0$, and by-eye disturbance, respectively. To 
produce by-eye ranking, we averaged the disturbance scores (1 to 3) obtained 
from all nine human experts.  We then ranked clusters by average disturbance 
score.

\bigskip
\section{Conclusions and future work}

In this work, we introduced a new cluster substructure statistic -- photon 
asymmetry ($\rm{A_{phot}}$), that measures the uniformity of the angular X-ray 
photon distribution in radial annuli. We compared photon asymmetry to two other 
measures of cluster morphology, power ratios (with a novel method for background 
correction) and centroid shifts, on the 400d cluster sample, and on simulated 
observations derived from it. Our focus was on performance of these substructure 
statistics in the low S/N regime, that is typical for observations of distant 
clusters.  Our main conclusions are as follows

\begin{enumerate}[itemsep=0pt]
  \renewcommand{\labelitemi}{$-$}
        
    \item The angular resolution of a cluster observation is far more important 
        than total counts for the ability to detect and quantify the 
        substructure.

    \item Both centroid shifts and photon asymmetry are significantly more 
sensitive to the amount of substructure than power ratios.

    \item Both centroid shifts and photon asymmetry agree well with by-eye 
classification.

    \item Centroid shifts are the best-performing substructure statistic in the 
        low spatial resolution ($\theta \gtrsim 5^{\prime\prime}$) and 
        background-dominated ($\sqrt{F_{bg}} \gtrsim F_{source}$) regimes. 
        
    \item Photon asymmetry is the best-performing substructure statistic in the 
low-counts regime.

    \item Photon asymmetry is the most sensitive measure of the \emph{presence} 
of substructure; 27 out of 36 clusters in the sample are classified by photon 
asymmetry as clusters with significant substructure (i.e., they are inconsistent 
with being axisymmetric), whereas the second best statistic, centroid shifts, 
finds significant substructure in only 21 out of 36 clusters.

    \item Photon asymmetry is the only statistics that is insensitive to 
    observational S/N below $\sim$ 1000 counts. Consequently, it is the only 
    statistic suitable for comparison of clusters and cluster samples across 
    large range of S/N, counts, backgrounds and redshifts. It is the best 
    candidate for studying the influence of substructure on bias and scatter in 
    scaling relations.

\end{enumerate}

We also suggested using concentration (a measure of cool core strength) and 
asymmetry (which quantifies merging or disturbance) as the main parameters for 
cluster classification. We find that clusters can demonstrate either a high 
degree of concentration or asymmetry, but not both at the same time. It is 
possible to use centroid shifts instead of photon asymmetry as the measure of 
cluster disturbance, but asymmetry is preferable given its better stability with 
respect to  observational S/N.

We are currently applying the photon asymmetry metric in a comparison of X-ray  
and SZ-selected cluster samples, to study the impact of morphology on cluster 
scaling relations, and to measure how morphology evolves with redshift.

\section*{Acknowledgments} \noindent{}D. N. acknowledges support by the 
National Science Foundation grant AST-1009012. M. M. acknowledges support by 
NASA through a Hubble Fellowship grant HST-HF51308.01-A awarded by the Space 
Telescope Science Institute, which is operated by the Association of 
Universities for Research in Astronomy, Inc., for NASA, under contract NAS 
5-26555. B.B. acknowledges support by NASA through Chandra Award Numbers 
13800883 issued by the Chandra X-ray Observatory Center, which is operated by 
the Smithsonian Astrophysical Observatory for and on behalf of NASA under 
contract NAS8-03060, and the National Science Foundation through grant 
ANT-0638937. E.M. acknowledges support from subcontract SV2-82023 by the 
Smithsonian Astrophysical Observatory, under NASA contract NAS8-03060.

We thank Christine Jones-Forman, William Forman, Marshall Bautz and Alastair 
Edge for their aid in morphologically classifying these galaxy clusters by eye.

\bibliographystyle{apj} 
\bibliography{library}

\clearpage
\begin{appendix}

    As explained in Section \ref{sec:morph_params}, our method of calculating 
asymmetry includes 2 steps: calculating the asymmetry in an annulus and 
combining the asymmetries from several annuli. To measure the asymmetry in each 
annulus we use the statistical framework of testing whether a given sample is 
drawn from a given probability distribution.  The sample in our case is the 
empirical angular photon distribution function $F_N$, and the given probability 
distribution is the true angular photon distribution function $G$ that would be 
produced by a perfectly circularly symmetric source. We note that $G$ is not 
trivial because of nonuniform detector illumination and various detector 
imperfections.

We define $F_N$ as the empirical cumulative angular distribution function of the 
photons in the $k$-th annulus:
\begin{equation}
    F_N(x) = \frac1N \sum_{R_{in}^k<r_i<R_{out}^k } {\textbf 1}\{\phi_i/2\pi \le 
x\},
\end{equation}
where ${\textbf 1\{A\}}$ is the indicator function of event $A$ and $N$ is the 
number of counts within the annulus $R_{in}^k<r<R_{out}^k$. Also, for 
convenience we rescale the angular range $[0,2\pi)$ to $[0,1)$.
Let $F$ be the true underlying distribution function for $F_N$, i.e $F$ is the 
limit of $F_N$ when $N \rightarrow \infty$.

Note that Kolmogorov-Smirnov, Cramer-von Mises and similar tests are usually 
used to check for the equality of 2 probability distributions. The values of 
these statistics give the probability of the null hypothesis (that the given 
sample is drawn from the given distribution), when compared to the null 
distribution. In our case, instead of checking whether $F_N$ is a realization of 
the known $F$ we need a measure of ``distance'' between $F$ and $G$ based on the 
measurement of $F_N$. In the following we show how one can use the value of 
Watson's test (a modification of Cramer-von Mises test suitable for 
distributions defined on a circle as opposed to a segment) to quantify the 
distance between $F$ and $G$ based on the sample $F_N$.

In the following we will use the notation
\begin{equation}
\begin{gathered}
 U^2[F,G;dH] = \int \bigg( F(x) - G(x) - \int \Big( F(x)-G(x) \Big) dH(x) \bigg)^2 dH(x),
\end{gathered}
\end{equation}
where $F$, $G$, and $H$ are arbitrary distribution functions defined on $[0,1]$, 
and all the integrals are taken over the same $[0,1]$ interval.

Using this notation, Watson's statistic $U_N^2$ is simply
\begin{equation}
U_N^2 = N \; U^2[F_N,F;dF].
\end{equation}

It can be viewed as a minimum of the $L_2$ distance between $F_N$ and $F$ over 
all possible points of origin on the circle \citep{Watson61}:
\begin{equation}
U_N^2 = \min_{\textrm{origin on the circle}} \int(F_N-F)^2dF.
 \end{equation}

In the limiting case $N \rightarrow \infty$, under the null hypothesis that the 
sample $\phi_i$ comes from the hypothesized distribution $F(x)$, the values of 
statistic $U_{\infty}^2 = \lim_{N \rightarrow \infty} U_N^2$ have the same 
distribution as $K^2 \pi^{-2}$, where $K$ is distributed according to 
Kolmogorov's distribution:
\begin{equation}
\begin{gathered}
    \textrm{Prob}\{U_{\infty}^2 < x\} = \textrm{Prob}\{K<\pi \sqrt{x} \} = 1 - 2 
\sum_{k=1}^{\infty} (-1)^{k-1}e^{-2k^2\pi^2 x^2}.
\end{gathered}
\end{equation}
We won't need the exact form of this limiting distribution, but we need to
know its mean which can be derived from known moments of Kolmogorov's 
distribution:
\begin{equation}
\langle U_{\infty}^2 \rangle = \frac{\langle K^2 \rangle}{\pi^2} =  
\frac{1}{12}.
\label{eq:U_N-mean}
\end{equation}

It may be shown \citep{Watson61} that given a discrete sample $x_1, x_2, \cdots, 
x_N$ hypothetically distributed according to $F(x)$, the statistic can be 
computed as
\begin{equation}
U_N^2 = \frac{1}{12N} + \sum_{i=0}^{N-1} \left(\frac{2i+1}{2N}-F_i\right)^2 - N 
\left(\frac12 - \frac1N \sum_{i=0}^{N-1} F_i\right)^2,
\end{equation}
where $F_i = F(x_i)$.

Now let's apply Watson's test statistic to the empirical distribution function 
$F_N$ and an arbitrary distribution function $G$ to which we need to compute a 
distance (in our method $G$ is the distribution function that represents a 
circularly symmetric source)
\begin{equation}
W_N^2 = N U^2[F_N,G;dG].
\label{eq:WN}
\end{equation}

Integrating by parts one can show that 
\begin{equation}
U^2[F_N,G;dG] = U^2[F_N,G;dF_N].
\label{eq:sym}
\end{equation}

Now we will replace $dF_N$ with $dF$ in the right hand side of \eqref{eq:sym}.  
While it is evident \citep{Doob1949} that as $N \rightarrow \infty$
\begin{equation}
U^2[F_N,G;dF_N] - U^2[F_N,G;dF] = R_N^2 \rightarrow 0 \textrm{ in probability,}
\label{eq:R2}
\end{equation}
the merit of this approximation and the rate of convergence are discussed below.

$U^2[F_N,G;dF]$ can be transformed in the following way
\begin{equation}
\begin{gathered}
U^2[F_N,G;dF] = \int \bigg[ \Big(F_N - F - \int (F_N-F)dF \Big) + \Big(F - G - 
\int (F-G)dF \Big) \bigg]^2 dF \\ = U^2[F_N,F;dF] + \int \Big(F_N - F - \int 
(F_N-F)dF \Big) \cdot \Big(F - G - \int (F-G)dF \Big) dF + U^2[F,G;dF]  
\label{eq:big}
\end{gathered}
\end{equation}

The first term, $U^2[F_N,F;dF]$ is distributed according to Kolmogorov's 
distribution and its mean is $\frac{1}{12}$ (see Eq.~\eqref{eq:U_N-mean}).

The second term,
\begin{equation}
V = \int (F_N-F-\Delta) g(x) dx, \;\; g(x) = (F - G - \int (F-G)dF) F'(x)
\end{equation}
has zero mean, because it is a sum of integrals of a function which has zero 
expectation value at any point on the segment
\begin{equation}
\langle F_N(x) - F(x) \rangle = 0 \;\; \forall x: 0<x<1
\end{equation}
with bounded functions $g(x)$ and $\int g(x) dx = const$.

The third term is the desired distance $d$ between $F$ and $G$.

Combining \eqref{eq:WN}, \eqref{eq:R2} and \eqref{eq:big} we find the following 
estimator $\hat{d}_N$ of $d$
\begin{equation}
\hat{d}_N = \frac{W_N^2}{N} - \frac{1}{12N}.
\label{eq:dhat}
\end{equation}
This estimator is biased by the average value of $R_N^2$.
\begin{equation}
\langle \hat{d}_N - d \rangle = \langle U^2[F_N,G;dF_N] - U^2[F_N,G;dF] \rangle 
= \langle R_N^2 \rangle   \end{equation}

We were not able to obtain an analytic bound on $R_N^2$ and its $N$-dependence.  
Judging by the form of \eqref{eq:R2},  $R_N^2$ should be of order $1/\sqrt{N}$.  
Considering this asymptotic behavior of $R_N^2$, our wish to explicitly correct 
for ``smaller'' bias $1/12N$ may look strange. The reason for this explicit 
correction is that $1/12N$ \emph{is bigger} than $R_N^2$ for relevant values of 
$N$ ($N<10^3$). We confirmed this statement by multiple numerical experiments 
with various distribution functions $F$ and $G$.  As $N$ reaches higher values 
($N \sim 10^4$) $\langle R_N^2 \rangle$ can become greater than $1/12N$, but 
both terms tend to zero with increasing $N$.

Now we need to take into account that the acquired light comes both from the 
cluster and the background. We model the counts distribution function $F$ as a 
weighted sum of cluster emission $F_{Cl}$ and a uniform background $G$ 
\begin{equation}
F = \alpha F_{Cl} + \beta G, \quad \alpha + \beta = 1, \quad \alpha = C/N,
\end{equation}
where $C$ is the number of cluster counts, and $N$ is the total number of counts 
in the given annulus. Then we obtain

\begin{equation}
    d = \int \Big(F-G - \int(F-G)dG\Big)^2 dG = \alpha^2 \int \Big(F_{Cl}-G - 
    \int(F_{Cl}-G)dG\Big)^2 dG = \alpha^2 d_{Cl}
\end{equation}

Now, using \eqref{eq:dhat} we see that 
\begin{equation}
    d_{N, Cl} = \frac{N}{C^2} \Big( W_N^2 - \frac{1}{12}\Big)
\end{equation}
is our estimator of the distance between the observed photon distribution  
function $F_N$ and the underlying cluster emission distribution function 
$F_{Cl}$.
 
The sum of distances $d_{Cl}^{(k)}$ in 4 annuli, where $k$ numbers the annuli, 
weighted by the estimated number of cluster counts in these annuli $C_k$, and 
multiplied by 100 gives \emph{photon asymmetry}:
\begin{equation}
    A_{phot} = 100 \sum_{k=1}^4 C_k d_{N_k, Cl}^{(k)} / \sum_{k=1}^4 C_k.
\end{equation}

 \end{appendix}

\input{galleries.tex}

\end{document}

%% file: six_authors.tex
\altaffiltext{\Harvard}{Department of Physics, Harvard University, 17 Oxford 
Street, Cambridge, MA 02138}
\altaffiltext{\MIT}{MIT Kavli Institute for Astrophysics and Space
Research, Massachusetts Institute of Technology, 77 Massachusetts Avenue,
Cambridge, MA 02139}
\altaffiltext{\CfA}{Harvard-Smithsonian Center for Astrophysics,
60 Garden Street, Cambridge, MA 02138}
\altaffiltext{\KICP}{Kavli Institute for Cosmological Physics, University of Chicago, 5640 South Ellis Avenue, Chicago, IL 60637}
\altaffiltext{\EFI}{Enrico Fermi Institute, University of Chicago, 5640 South Ellis Avenue, Chicago, IL 60637}

\def\Harvard{1}
\def\MIT{2}
\def\CfA{3}
\def\KICP{4}
\def\EFI{5}

\author{
D.~Nurgaliev\altaffilmark{\Harvard},    
M.~McDonald\altaffilmark{\MIT},
B.~A.~Benson\altaffilmark{\KICP,\EFI}
E.~D.~Miller\altaffilmark{\MIT},
C.~W.~Stubbs\altaffilmark{\CfA,\Harvard}, 
A.~Vikhlinin\altaffilmark{\CfA},
}

%% file: galleries.tex
\begin{turnpage}
\begin{figure*}[p]
    \begin{center}
        \epsscale{1.17}
        \plottwo{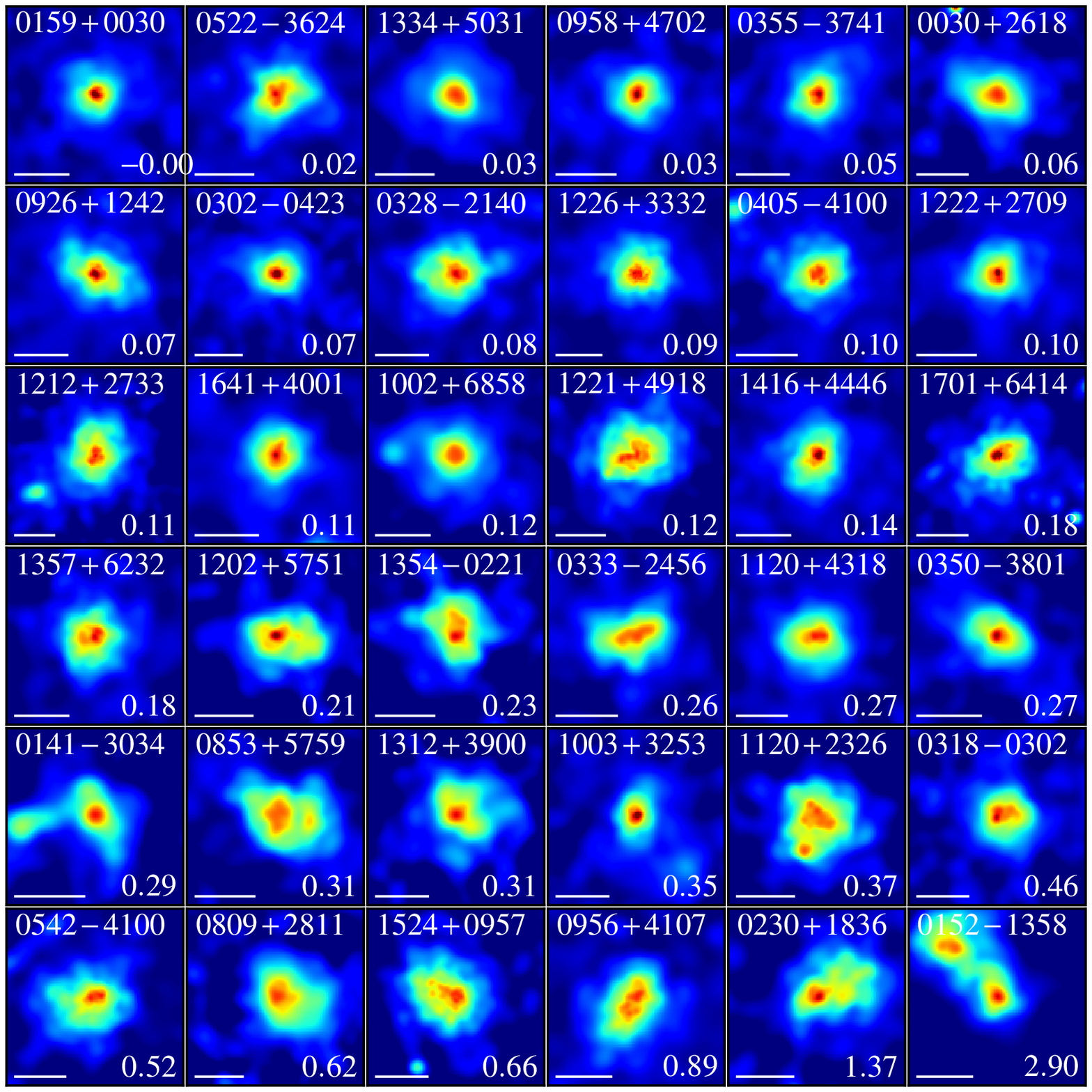}{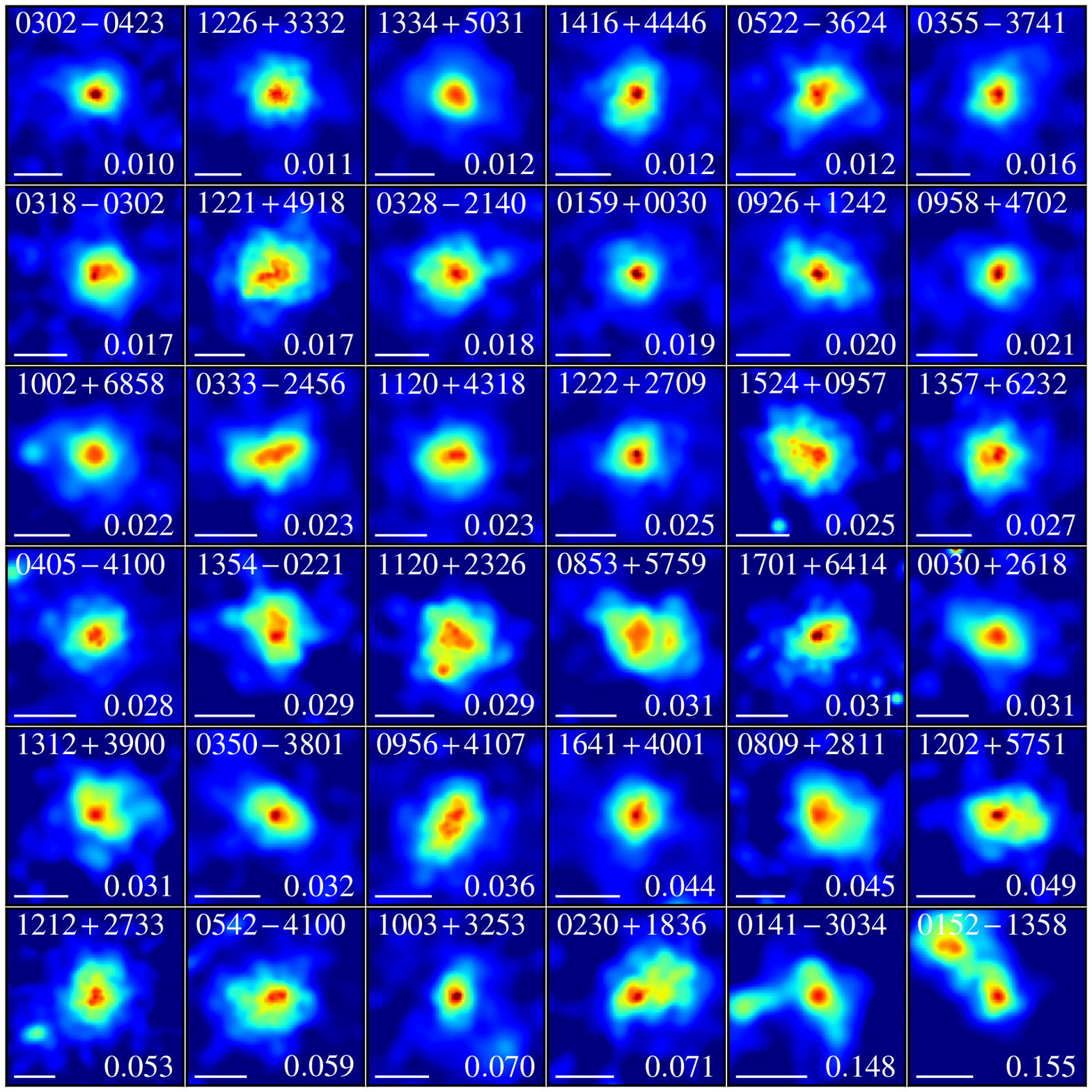}
        \caption{
            Clusters sorted by asymmetry (left) and centroid shifts (right).  
            The value of the substrucure statistics increases top-to-bottom, 
            left-to-right in both plots. Cluster name is in upper right corner, 
            the value of the statistic is in the upper left corner.  The names 
            of the clusters are identical to those used in \citet{CCCPII}.
            \emph{Left plot}. Clusters sorted by the value of asymmetry - the 
            new substructure measure which is presented in this paper.
            \emph{Right plot}. Clusters sorted by the value of centroid shifts.
        }
        \label{fig:gal-asym-w}
    \end{center}
\end{figure*}
\end{turnpage}

\begin{turnpage}
\begin{figure*}[p]
    \begin{center}
        \epsscale{1.16}
        \plottwo{gal_asym.eps}{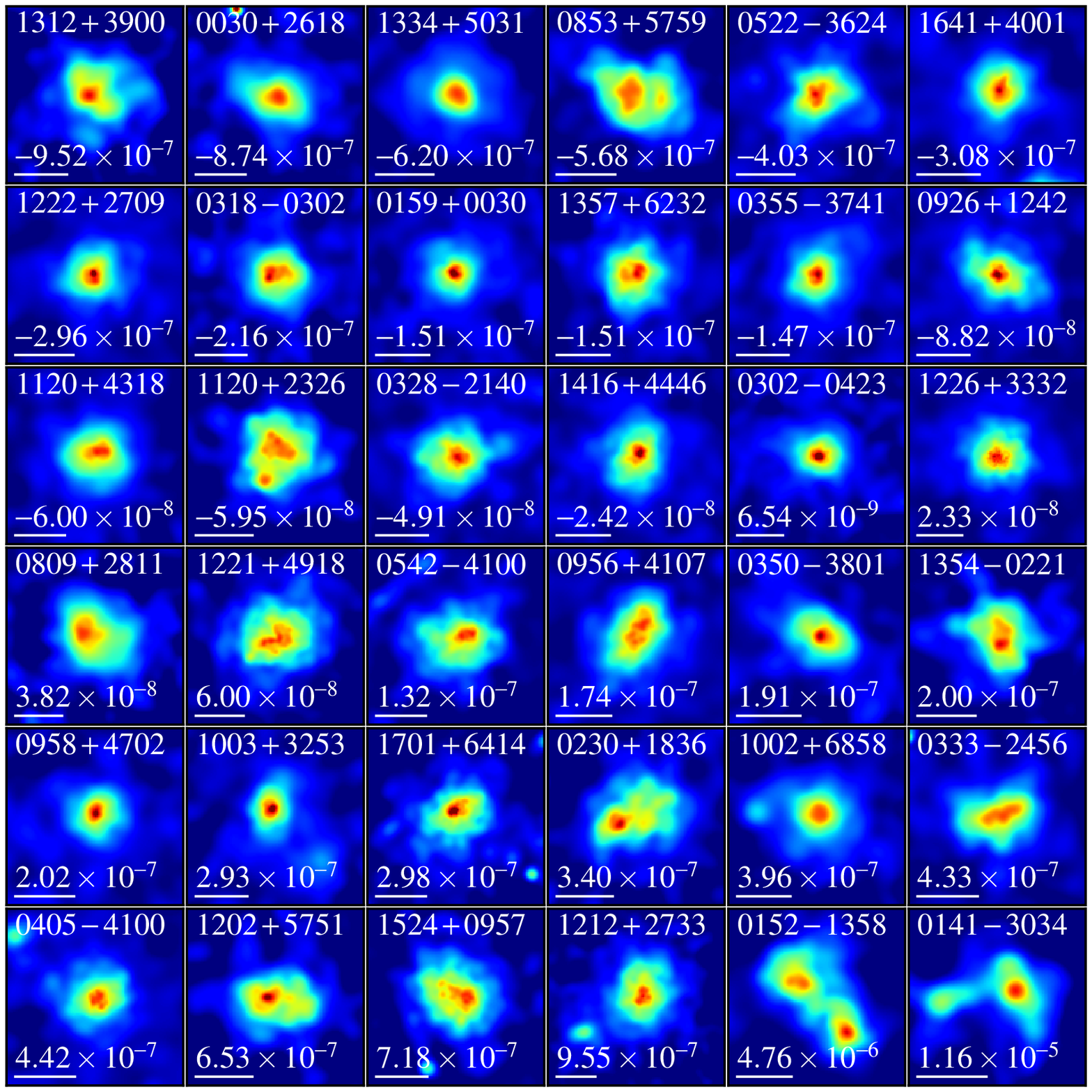}
        \caption{
            Clusters sorted by asymmetry (left) and power ratios $P_3/P_0$ 
            (right). The value of the substrucure statistics increases 
            top-to-bottom, left-to-right in both plots. Cluster name is in upper 
            right corner, the value of the statistic is in the upper left corner  
            The names of the clusters are identical to those used in 
            \citet{CCCPII}.
            \emph{Left plot}. Clusters sorted by the value of asymmetry - the 
            new substructure measure which is presented in this paper.
            \emph{Right plot}. Clusters sorted by the value of $P_3/P_0$.  White 
            circle is the aperture used for calculating power ratio ($R_{500}$).
        }
        \label{fig:gal-asym-P30}
    \end{center}
\end{figure*}
\end{turnpage}

\begin{turnpage}
\begin{figure*}[p]
    \begin{center}
        \epsscale{1.17}
        \plottwo{gal_asym.eps}{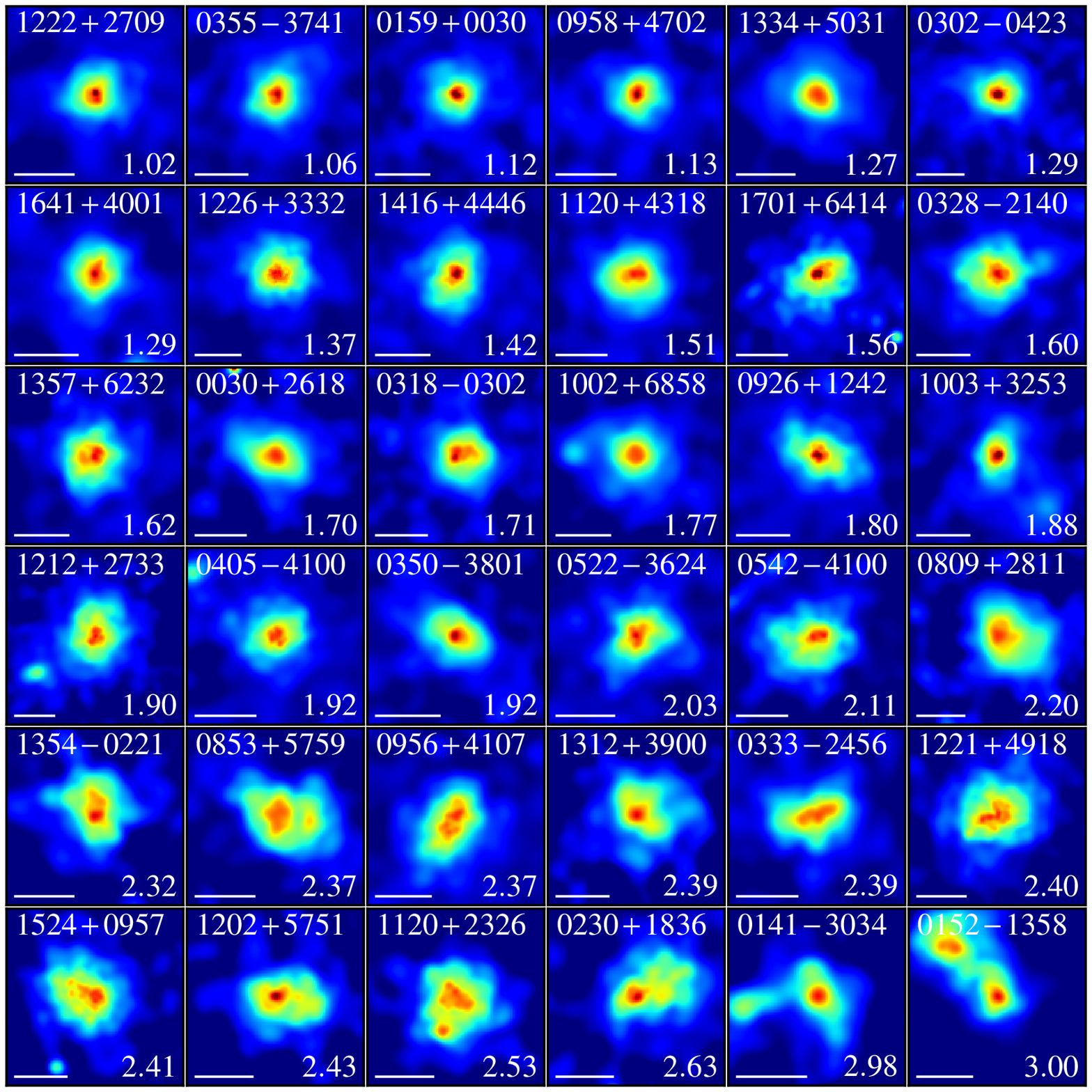}
        \caption{
            Clusters sorted by asymmetry (left) and "by-eye" level of 
            disturbance (right).  The value of the substrucure statistics 
            increases top-to-bottom, left-to-right in both plots.  Cluster name 
            is in upper right corner, the value of the statistic is in the upper 
            left corner  The names of the clusters are identical to those used 
            in \citet{CCCPII}.
            \emph{Left plot}. Clusters sorted by the value of asymmetry - the 
            new substructure measure which is presented in this paper.
            \emph{Right plot}. Clusters sorted by the average value of their 
            "disturbness" evaluated by 4 numan experts.  }
        \label{fig:gal-asym-byeye}
    \end{center}
\end{figure*}
\end{turnpage}